\documentclass[aps,reprint,superscriptaddress]{revtex4-2}
\usepackage{graphicx}% Include figure files
\usepackage{dcolumn}% Align table columns on decimal point
\usepackage{bm}% bold math
\usepackage{comment}
\usepackage{float}
\usepackage{graphicx}
\usepackage{titlesec}
\usepackage{lmodern}
\usepackage{subfiles}
\usepackage{standalone}
\usepackage{xcolor}
\usepackage{soul}

\begin{document}

\preprint{APS/123-QED}

\title{\textbf{Ultra High-Q tunable microring resonators enabled by slow light} 
%\thanks{A footnote to the article title}
}% 

\author{Priyash Barya}
\affiliation{Department of Electrical and Computer Engineering, University of Illinois at Urbana-Champaign, Urbana, IL 61801 USA}

\author{Ashwith Prabhu}
\affiliation{Department of Physics, University of Illinois at Urbana-Champaign, Urbana, IL 61801 USA}

\author{Laura Heller}
\affiliation{Department of Electrical and Computer Engineering, University of Illinois at Urbana-Champaign, Urbana, IL 61801 USA}

\author{Edmond Chow}
\affiliation{Holonyak Micro and Nanotechnology Laboratory, University of Illinois at Urbana-Champaign, Urbana, IL 61801 USA}

\author{Elizabeth A. Goldschmidt}
\altaffiliation{ goldschm@illinois.edu}
\affiliation{Department of Physics, University of Illinois at Urbana-Champaign, Urbana, IL 61801 USA}

\begin{abstract}
High-Q nanophotonic resonators are crucial for many applications in classical and quantum optical processing, communication, sensing, and more. We achieve ultra-high quality factors by preparing a highly transparent and strongly dispersive medium within a resonator, causing a reduction in the group velocity that leads to a nearly three order of magnitude increase in the quality factor. We implement this via spectral hole burning in erbium-doped thin-film lithium niobate microring resonators, and show Q-factors exceeding $10^8$.  Additionally, we show that the interplay between the spectrally narrowed resonance and the broader bare resonance produces a Fano lineshape, which we dynamically control via electro-optic tuning. Finally, we present a theoretical model for our experimentally observed resonator linewidths, which are not well-described by the standard Bloch equations. Our results show a dramatic reduction in the erbium dephasing rate under a strong optical drive, leading to much narrower linewidths than would otherwise be expected given the large circulating intensity in the resonator.  %

\end{abstract}

\maketitle

High-Q optical resonators are essential components in nanophotonics, enabling narrowband filtering \cite{zhao2024cavity}, enhanced nonlinearity \cite{lu2020toward, holzgrafe2020cavity},  cavity quantum electrodynamics \cite{walther2006cavity, fan2018superconducting}, sensing \cite{armani2007label}, precision spectroscopy \cite{dutt2019experimental,dutt2018chip},  laser stabilization \cite{hafner20158}, and more. Quality factors in these resonators are fundamentally limited by optical losses from material absorption and surface scattering, and can be as high as $10^7$ for microring resonators in the thin-film lithium niobate platform studied here \cite{zhu2024twenty, zhang2017monolithic}. Higher quality factors would have a direct impact for enabling functionality on chip that is currently only possible in macroscopic systems, which is vital for making scalable, robust, and deployable devices. For instance, the most stringent sensing and spectroscopic applications require Q-factors that are far out of reach for nanophotonics and only possible in macroscopic cavities ($Q\gtrsim10^{11}$, \cite{ludlow2007compact, hafner20158}) and atomic systems ($Q\gtrsim10^{15}$, \cite{boyd2006optical,schioppo2017ultrastable,PhysRevResearch.3.023152}). Higher Q factors would also enable new functionalities for building scalable quantum networks such as integrating ultra-narrowband photon generation and efficient quantum memory on-chip. Continued materials and fabrication improvements will likely lead to continued incremental improvements in Q, but here we pursue a pathway that gets around the materials and fabrication limits by engineering a slow group velocity for light propagating in the cavity. This approach uses the fact that optical losses are propagation losses, and therefore depend on the propagation length, while Q depends on the rate of optical loss in time. Thus, a reduction in the group velocity of light leads directly to an increase in the Q by the same factor. High optical depth atomic ensembles enable group velocity reduction of many orders of magnitude via electromagnetically induced transparency \cite{hau1999light, lukin2001controlling} and other schemes \cite{bigelow2003superluminal, bigelow2003observation}, however, incorporating such ensembles with nanophotonics is challenging. Previous work on resonance narrowing via slow light has primarily focused on two regimes: very slow group velocities in macroscopic cavities \cite{sabooni2013spectral, huet2016millisecond, muller1997optical} and modest group velocity reduction via band engineering in nanophotonic cavities \cite{lu2022high, lee2011slow}. 

\begin{figure*}[hbt!]
\includegraphics[width=\linewidth,keepaspectratio]{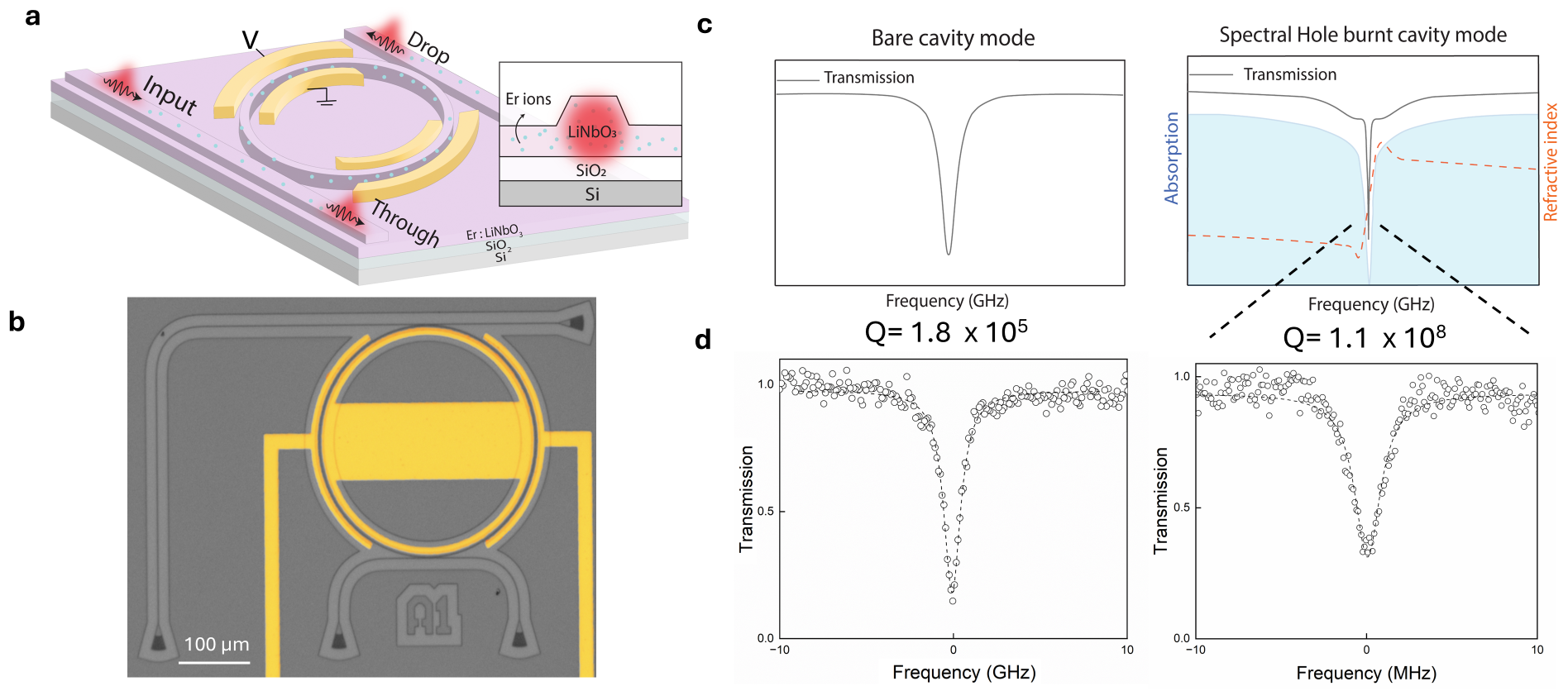}
\caption{\label{fig: MainFig} \textbf{Slow light induced linewidth narrowing.} (a) Schematic of an Er:TFLN ring resonator cavity coupled to through and drop ports with electro-optic tuning. Inset: schematic cross-sectional view of the waveguide.
(b) Optical image of the fabricated microring resonator, which was loaded onto a stage at a temperature of $\sim$ 2 K. The gold coloured regions show the electrodes utilized in the EO tuning. Light was coupled in and out using grating couplers and an additional drop port at the top is used for the slow light measurements.
(c) Schematics of a bare cavity
mode with no absorption (left) and ultra high-Q resonator mode (right) formed in a narrow transparency window in a region of high absorption that creates a highly dispersive medium with ultra-low group velocity. (d) Measured transmission through a resonance tuned away from the Er absorption (left) and a narrowed resonance (right) showing narrowing by nearly three orders of magnitude due to preparing a spectral hole in a region of high absorption.}
\end{figure*}

We use a unique platform, erbium-doped thin-film lithium niobate (Er:TFLN) \cite{yang2023controlling, wang2020incorporation}, in which a high optical depth ensemble of coherent Er emitters is directly coupled to low-loss nanophotonic structures fabricated in TFLN. We employ spectral hole-burning (SHB), a powerful technique for absorption and dispersion engineering in rare-earth ensembles \cite{dutta2019integrated, dutta2023atomic} and other media \cite{shakhmuratov2005slow, lauro2009slow}. A narrowband laser selectively depopulates or shelves a spectrally narrow subset of emitters from a large inhomogeneously broadened ensemble, thereby creating a narrow transparency window (and a corresponding sharp slope in the index of refraction) that can persist for ms or longer depending on the medium. The group velocity of light propagating through this spectral hole $v_g = c/(n(\omega)+\omega\partial n/\partial\omega)\approx\Delta_{\text{hole}}/\alpha$ where $\Delta_{\text{hole}}/(2\pi)$ is the width of the spectral hole and $\alpha$ is the absorption coefficient outside the spectral hole \cite{lauro2009slow, sabooni2013spectral,shakhmuratov2005slow}. This can be dramatically reduced from $c$ for typical rare-earth spectral holes with $\Delta_{\text{hole}}/(2\pi)\gtrsim\rm{MHz}$ and $\alpha\gtrsim\rm{cm}^{-1}$. In this work, we show a nearly three order of magnitude decrease in the group velocity, and a corresponding Q-factor greater than $10^8$, beyond the state of the art for TFLN to date. Such narrow resonances on chip enable new functionality for tunable spectral filtering \cite{xu2005micrometre}, photon storage \cite{zhang2019electronically}, laser stabilization \cite{liu2022photonic}, and more \cite{yuan2021synthetic, dutt2022creating}. 

We also work out a theoretical model for the unique regime we are able to study in the Er:TFLN platform. We show that the narrowing can be understood as an interference of resonances, which leads to tunable Fano lineshapes, which we demonstrate by electro-optic (EO) tuning of our narrowed Fano feature. We further model the spectral hole width and show that the large circulating intensity in the resonator during the hole-burning phase decouples the Er emitters from the surrounding spin bath and dramatically reduces the excess dephasing. This results in much narrower features than expected based on a naive Bloch equation treatment. We showcase the necessary modification to the Bloch equations and show good agreement with our data, a vital step for accurately modeling and understanding spectral hole-burned features in rare-earth emitters coupled to resonators.

\subsection{Slow light-induced linewidth narrowing}
The maximum achievable Q-factor ($Q = \omega_c/\kappa$) of resonators is primarily constrained by intrinsic loss mechanisms ($\kappa > \kappa_{i}$), which includes both the material absorption rate and imperfect mode confinement, resulting in coupling to undesired radiation modes. We overcome this limitation by adding a highly dispersive medium within the resonator, which leads to a significant reduction in the group velocity of light ($v_g$). When introduced into a microcavity, this reduction increases the round-trip time of circulating light, effectively extending the cavity lifetime and enhancing the Q-factor. One might initially guess that such a slowdown would actually increase the coupling to lossy modes, as photons would interact with the absorptive medium and radiative decay channels for a longer time. However, Soljačić et al. \cite{soljavcic2005enhancement},  demonstrated that dispersion not only reduces the group velocity but also slows down the coupling rates to lossy modes, thereby increasing the Q-factor,  which scales as  $Q=\frac{c}{nv_g}(\frac{\omega_c}{\kappa})$, where $n$ is the group index without a tranperancy window, and $\omega_c$ is the cavity center in angular frequency. 

\begin{figure}
\includegraphics[width=\linewidth,keepaspectratio]{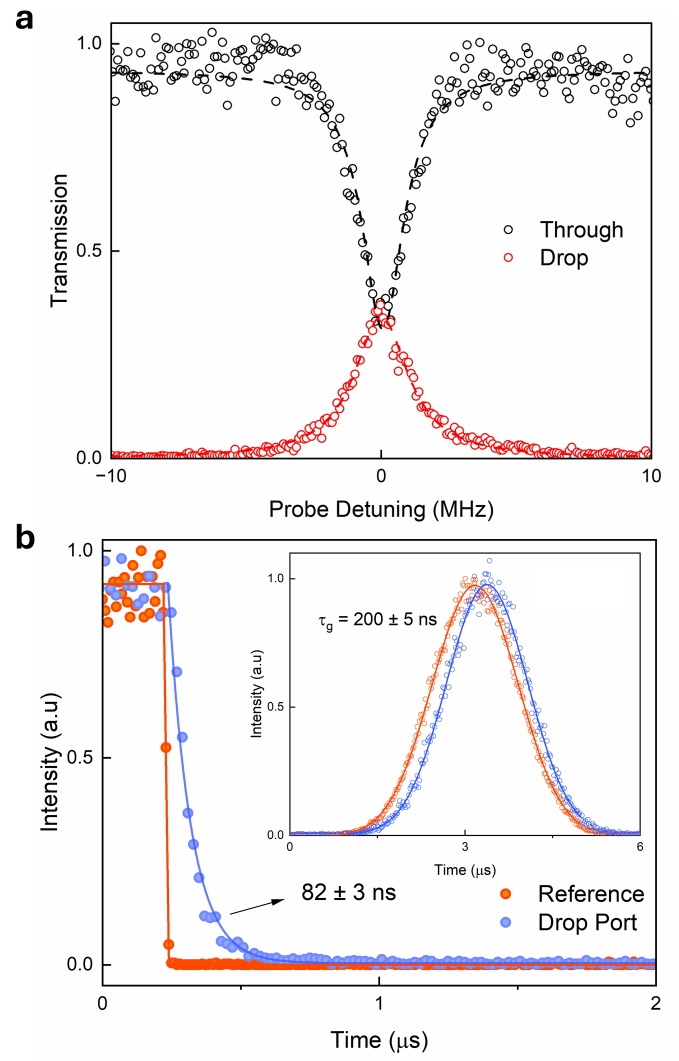}
\caption{\label{fig:CavLife}  \textbf{Extending Cavity Lifetime.} (a) The narrowed resonance measured at the through port (black) and the drop port (red)(b) The extended ring down time of the cavity measured through the drop port (blue), showcasing a significantly longer time when compared with the bare cavity mode (red) measured at a resonance off the absorption. Inset: delay of a narrowband gaussian pulse measured through the drop port.}
\end{figure}
In our system, we first investigate the bare cavity resonance of the microring resonators away from the Er absorption band. Two bus waveguides are used to evanescently couple light in and out of the ring cavity with a rate of $\kappa_{\text{ext}} = 2\pi \times324 \text{ MHz}$ each (Supplemental), while the intrinsic loss rate was extracted to be $\kappa_{i}= 2\pi \times471 \text{ MHz}$, resulting in the total  $\kappa = 2\pi \times1.12 \text{ GHz}$ or a Q-factor of $1.8\times10^5$ (Fig. \ref{fig: MainFig}d). The 300~$\mu\rm{m}$ diameter of our resonators means we are guaranteed at least 2 resonances within the $\sim$ 158 GHz wide Er absorption band on the \textit{I$_{15/2}$}-\textit{I$_{13/2}$}  transition centered at 1531.6~nm. The large inhomogeneous linewidth is due to charge compensation and disorder created when Er substitutes for lithium, as well as the intrinsic disorder in congruent lithium niobate. We use devices with heavy erbium doping (0.5\%) to achieve high optical depth. The resulting Er absorption introduces an additional loss rate of  $\kappa_{\rm{Er}} = 2\pi \times 16$ GHz (Supplemental), which dominates both intrinsic and coupling losses, leading to significantly broadened and undercoupled resonances within the erbium absorption bandwidth.  We next perform SHB at the center of the cavity resonance by optically pumping a spectrally narrow subset Er ions. 

This results in a narrow transparent window, or spectral hole, in the absorption profile at the frequency of the burn pulse. The width of the spectral hole is fundamentally limited to be larger than twice the measured homogeneous linewidth, $\Delta_{\text{hole}}/2\pi \ge  490 \text{ kHz}$ (Supplemental). However, due to power broadening the holes are much broader, estimated to be between 30-100 MHz for the range of optical powers used here (Supplemental), which is still much narrower than the resonator $\kappa$ even away from the Er absorption. The narrow transparency window is accompanied by strong dispersion, resulting in a dramatically reduced group velocity and a longer time for the photons to finish each round trip, described by the expression $nv_g/c = \Delta_{\text{hole}}/\kappa_{\rm{Er}}$. Consequently, this results in a narrowed resonance mode at the frequency of the spectral hole. Since the decay rate of the spectral hole is slow compared to all other frequency scales ($<10~\rm{kHz}$), we can ignore any hole dynamics beyond the approximately flat spontaneous emission background we observe during our measurements. We note that moving to lower temperatures and/or higher magnetic fields would extend the lifetime of our spectral holes beyond the Er optical lifetime \cite{thiel2010optical, askarani2020persistent} and enable operation without this spontaneous emission background. We measure the transmission as a function of detuning by sending a weak probe pulse $4~\mu\rm{s}$ after the burn pulse and detecting at the through port of the device. We observe a sharp Lorentzian dip at the burn frequency with a width of 1.9 MHz, corresponding to a Q-factor of $1.1 \times 10^8$ (Fig. \ref{fig:CavLife}a). Further, we measured the output at the drop port and observed a corresponding peak, consistent with the microring resonator model. 

To verify that the Q-factor extends the cavity lifetime, we perform time-domain measurements to measure the ringdown time. Specifically, we send an optical square pulse and measure the decaying tail at the falling edge at the drop port (Methods). Our measurements reveal that the narrowed cavity mode exhibits a lifetime of 82 ns, whereas the bare cavity (in the absence of absorption) has an expected lifetime of 0.14 ns, below the resolution of our detection scheme (Fig. \ref{fig:CavLife}b). This significant increase further validates the formation of a high-Q resonance mode and confirms the extended cavity lifetime.

To further investigate the slow light effect, we measure the resonant group delay experienced by a Gaussian pulse. We send a narrowband, 1.5~$\mu$s Gaussian pulse through the resonance and observe a 200~ns delay compared to the bare cavity mode case (Fig. \ref{fig:CavLife}b inset). This corresponds to a group velocity reduced to nearly $2 \times 10^5$ m/s in the ring waveguide, $\sim 700\times$ slower than the speed of light in the LN waveguide. This result aligns closely with our theoretical estimates (Eq. S41) as well as the scaling of the Q-factor compared to the bare cavity Q. The delay could be further extended by employing a coupled resonator optical waveguide (CROW) architecture, which can provide a linear scaling of pulse delay with the number of coupled rings \cite{bogaerts2012silicon}. The current delays represent some of the longest optical pulse delays achievable on an integrated chip, offering promising opportunities for scalable quantum memory and quantum information applications.

\begin{figure}
\includegraphics[width=\linewidth,keepaspectratio]{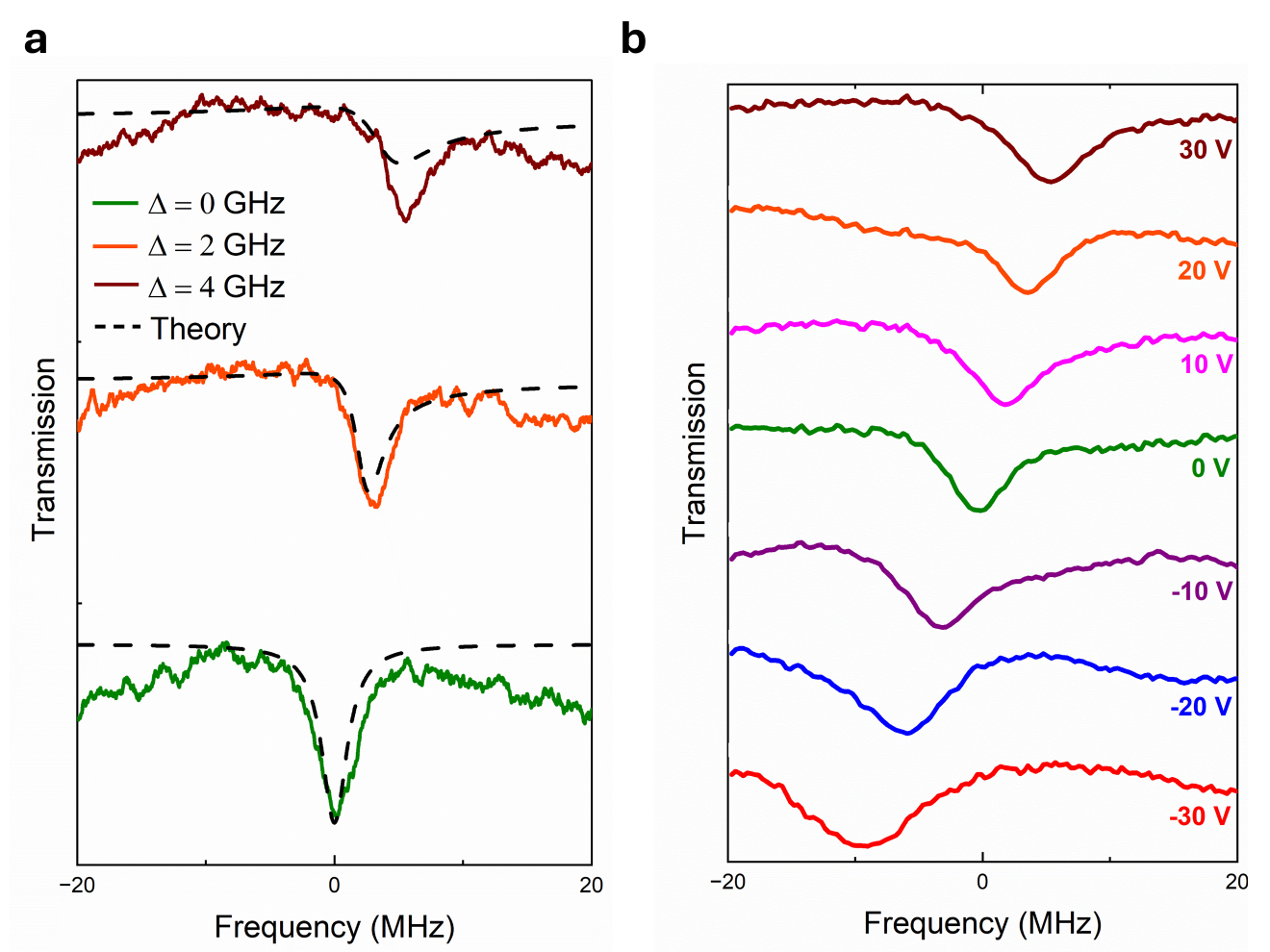}
\caption{\label{fig:Fano} \textbf{Fano Lineshape and EO tuning.}
(a) Experimental data and theoretical model showing the Fano lineshape at different laser detunings from the bare cavity resonance. (b) Here, the bare cavity resonance is tuned electro-optically while keeping the burn frequency constant, demonstrating a tunable Fano filter with dynamic control over the resonance characteristics. }
\end{figure}

\subsection{Fano Lineshape and Electro-optic tuning} 
Fano-like resonances in photonics arise from the interference between a spectrally narrow resonant mode and a broader continuum \cite{limonov2017fano}. In our system, the linewidth-narrowed resonator mode—induced by the slow-light effect—acts as the narrow resonance, while the damped bare cavity transmission serves as the continuum. As a result, the relative detuning between the angular frequency of the burn laser ($\omega_l$) and the bare cavity mode ($\omega_c$) determines the interference phase, giving rise to the characteristic asymmetric Fano lineshape. 

This behavior can be rigorously derived from our theoretical treatment (Supplemental), which showcases the classic asymmetric lineshape (Fig. \ref{fig:Fano}a) and also naturally reveals the presence of two distinct resonator modes. This observation further supports our interpretation that the narrow resonance behaves as a separate resonant mode. By comparing our model to the standard Fano resonance equation ($T \propto \frac{(q + \Omega')^2}{1 + \Omega'^2}$) \cite{limonov2017fano}, we can extract the normalized detuning ($\Omega'$) and the Fano asymmetry parameter as:

\begin{equation}
\label{eq:Fano}
q =  \frac{\omega_c - \omega_l}{ \frac{\kappa}{2} + \frac{\kappa_{\text{Er}}}{2}}
\end{equation}

To experimentally validate this, we detune the burn laser frequency with respect to the cavity mode. This resulted in a progressively negative asymmetry parameter ($q$), shifting the resonance profile from a quasi-Lorentzian to a distinctly asymmetric Fano shape. The spectral dip was observed to undergo a blue shift, consistent with the well-known Fano response in photonics.

In addition to tuning the burn laser frequency, we can also manipulate the bare cavity mode itself while keeping the laser frequency fixed to achieve the asymmetric resonance shape. The high $\chi^{(2)}$ nonlinearity of lithium niobate (LN) provides an effective mechanism for this control. By applying a DC voltage across the resonator, we electro-optically shift the bare cavity resonance frequency by a rate of 0.12 GHz/V (Supplemental), thereby tuning the Fano lineshape. As shown in Fig. \ref{fig:Fano}b, this voltage-induced tuning effectively shifts the spectral dip of the Fano resonance, offering dynamic reconfigurability.

These high-Q resonator modes exhibit some of the narrowest filter responses demonstrated on an integrated photonic platform. Combined with the fast electro-optic tunability of LN ($>$GHz), this approach enables a wide range of applications in dynamically reconfigurable filtering, high-speed photonic switching, and adaptive on-chip laser stabilization.

\subsection{Modified Bloch Equation model}

\begin{figure}[H]
\includegraphics[width=\linewidth,keepaspectratio]{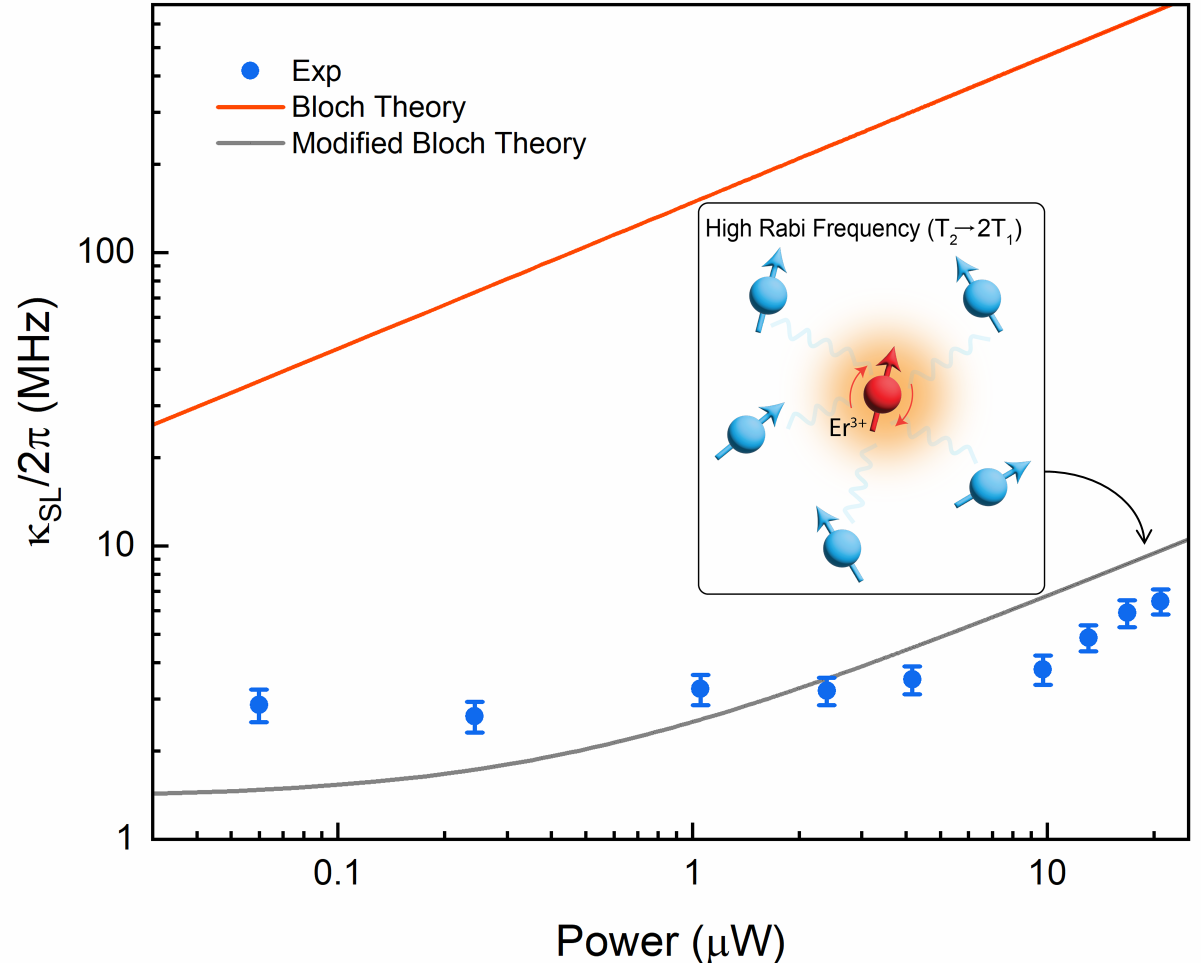}
\caption{\label{fig:Model} \textbf{Modified Bloch Equation model.} The power ($P_{in}$)-dependent linewidth of the resonator at zero detuning from the bare cavity resonance. The data (blue points) showcases a stark deviation from the standard Bloch theory (red line). The modified Bloch theory (grey line) matches the data much better, indicating a lower decoherence rate at higher power. The inset illustrates the physical mechanism behind the approach, where the emitter is decoupled from the spin bath due to the large Rabi frequencies in the cavity.}
\end{figure}

We next examine the dependence of resonance linewidth on input pump power ($P_{\text{in}}$), as illustrated in Fig. \ref{fig:Model}. To provide deeper insight, our model diverges from heuristic group velocity approach used earlier and instead considers a more fundamental approach of analyzing a cavity coupled to an atomic ensemble (Supplemental) \cite{lei2023many}. One might expect the standard optical Bloch equations, incorporating experimentally extracted system parameters including the spontaneous decay rate ($\gamma_s$), decoherence rate ($\gamma$), cavity loss rates ($\kappa_i, \kappa_{\text{ext}}, \kappa_{\rm{Er}}$), and the single-photon coupling rate ($g$) to explain the resonance width (see Supplemental for details). In this model the slow light induced narrowed resonance loss rate ($\kappa_{\text{SL}}$) would be expected to follow the form:

\begin{equation}
\label{eq: Narrowing}
\kappa_{\text{SL}} =  \left( \frac{\Delta_{\text{hole}}}{\kappa_{\text{Er}}} \right) \kappa
\end{equation}

 where the spectral hole can be expressed as $\Delta_{\text{hole}} = 2\Omega\sqrt{\frac{\gamma}{\gamma_s}}$, including the Rabi frequency given by $\Omega =  \frac{2 g \sqrt{\kappa_{\text{ext}}} }{ \kappa / 2} \sqrt{\frac{P_{\text{in}}}{\hbar \omega_c}}$. Since, the group velocity can be expressed as $nv_g/c = \Delta_{\text{hole}}/\kappa_{\rm{Er}}$, the theory naturally reveals the scaling of the cavity linewidth ($\kappa$) with the group velocity.

The simple Bloch model exhibits a striking discrepancy with the experimentally observed linewidths (Fig. \ref{fig:Model}), which are up to two orders of magnitude narrower than the theoretical prediction. This deviation cannot be attributed solely to experimental uncertainties in parameter extraction (see Supplemental) and, furthermore, the observed linewidths do not follow the square root dependence on optical power predicted by the Bloch model.

The failure of the conventional Bloch equations stems from the underlying assumption that the Lindbladian describing the dephasing and decoherence is independent of the optical field strength. However, prior investigations of rare-earth-doped solids such as Pr$^{3+}$: LaF$_3$ have demonstrated that the dephasing induced by a dynamic bath is effectively suppressed when the Rabi oscillation frequency at high intensity exceeds the correlation time $\tau_c$ of the bath \cite{devoe1983experimental}.
 
In our system, the erbium ions couple to a dense nuclear spin bath composed of lithium and niobium nuclear spins as well electron spins of other erbium ions, a phonon bath, and an electronic bath consisting of charge traps and charged two level systems. Irrespective of the microscopic details of the bath, it can modeled as stochastic process  with a correlation time $\tau_c$. The bath and, consequently, the corresponding stochastic process, can be assumed to be Markovian when the condition $\sqrt{\gamma\tau_c}< 1$ is satisfied. At an intuitive level, fast Rabi flopping results in time averaging of the dephasing, similar to continuous dynamical decoupling.

Reaching the requisite Rabi frequencies to observe this effect in bulk rare-earth-doped materials is difficult, as the optical intensity required is typically out of reach. However, in a nanophotonic waveguide with sub-micron mode area ($A_{\text{mode}} = 0.78~\mu\rm{m}^2$) the local intensity is greatly enhanced. Additionally, in a ring resonator configuration, the circulating optical power is further amplified. For instance, an input power  $P_{in} = 21 \ \mu \text{W}$ in the bus waveguide leads to a Rabi Frequency   $\Omega = 2\pi \times 106 \text{ MHz}$ in the ring resonator. To account for this phenomenon, we introduce a dephasing rate described by the modified Bloch equation formalism \cite{geva1995relaxation,berman1985modified,schenzle1984microscopic}, given by:

\begin{equation}
\label{eq:modified}
  \gamma_{\text{mod}} = \frac{\gamma}{1 + (\Omega\tau_c)^2} +\frac{\gamma_s}{2}  
\end{equation}

 At sufficiently high optical power ($\Omega\gg 1/\tau_c$) the dephasing rate asymptotically approaches the spontaneous emission limit ($\gamma_{\rm{mod}}\rightarrow\gamma_s/2$), thereby eliminating spin-bath-induced broadening. By incorporating \( \gamma_{\text{mod}} \) and fitting the experimental data, we extract a correlation time of $\tau_c \approx 1~\mu s$. This value lies well within the range of typical correlation times previously observed in Er:LN systems \cite{thiel2010optical} and satisfies the Markovianity condition $\sqrt{\gamma\tau_c}< 1$. With this correction, the theoretical prediction aligns closely with the experimental data, effectively resolving the discrepancy encountered with the conventional model. It is also important to note that the experimental data also aligns qualitatively with the trend predicted by the modified theory, deviating from the standard $\kappa_{SL} \propto \sqrt{P_{\mathrm{in}}}$ dependence expected from the Bloch model. This finding is consistent with prior studies \cite{berman1985modified,devoe1983experimental} demonstrating that optical linewidths in impurity-ion solids can exhibit a field-dependent narrowing due to suppression of local environmental fluctuations, but goes beyond past demonstrations that could not reach such high Rabi rates. This modified model is vital for understanding the limits on (and possibilities for) atomic frequency comb storage in nanophotonic devices as it dramatically changes the optical power required to prepare a high resolution comb in the presence of a dephasing spin bath.

\section{Discussion}
The slow-light narrowing we demonstrate here can be combined with narrower initial resonators to reach even higher Q factors, and can be used for a variety of classical and quantum photonic devices. It can also be extended to other rare-earth dopants with transitions in other frequency bands throughout the visible and near-infrared \cite{thiel2011rare}. Improving our fabrication to reach the state-of-the-art in TFLN ring resonators ($\rm{Q}=2\times10^7$ \cite{zhu2024twenty}) would mean slow light resonances with Q as high as $10^{10}$, corresponding to a cavity linewidth as small as $\kappa<10~\rm{kHz}$ and a ringdown time as long as $\tau>10~\mu\rm{s}$. This linewidth is smaller than most atomic transitions used for spin-photon interfaces. Such a system would be well suited for dynamically tunable on-chip filtering of photonic qubits from atomic spin qubits, noting that the photons would have to be either naturally matched to a rare-earth absorption band or undergo wavelength conversion. Furthermore, lithium niobate is a common platform for entangled photon pair generation, which is challenging to implement with sufficiently narrow linewidths to match atomic qubits. Tunable-width, slow-light resonator modes would enable photon pair generation with frequencies and linewidths engineered to match atomic systems.

The narrow linewidth and long cavity lifetime also can be used for many other applications that are currently out of reach on-chip. Currently, laser stabilization for the most stringent systems can only be done with bulk, $\gtrsim\rm{cm-scale}$ systems like macroscopic cavities \cite{boyd2024basic}, atomic ensembles with narrow spectral features \cite{olson2019ramsey}, and bulk crystals with narrow spectral holes \cite{thorpe2011frequency}. Recently, such stabilization was shown with a bulk analog of our system, a slow light resonance in a bulk cavity \cite{gustavsson2025using}. Our system enables such precise laser stabilization on chip, opening up many possibilities for chip-scale clocks and more compact laser systems for atom and ion traps for quantum computing. Finally, the ability to electro-optically modulate the TFLN platform, combined with a long cavity lifetime and engineerable dispersion, enables novel pulse shaping applications over MHz-scale frequency ranges, either in the telecom band with Er or in another rare-earth absorption band to match other qubit platforms.

Importantly, the system presented here has two major differences from a resonator with similarly high Q achieved without reducing the group velocity. One is the spontaneous emission noise that prevents operation at the single photon level. This can be eliminated by preparing persistent spectral holes, which has been demonstrated in erbium-doped lithium niobate \cite{askarani2020persistent, thiel2010optical} but is not possible in our apparatus due to restrictions on temperature and magnetic field. The other difference is more fundamental, the circulating electric field in the resonator does not increase between the bare resonance with no absorption and the slow light narrowed resonance. Slowing light via coupling to an atomic medium comes with a reduction in electric field that exactly matches the increase in the time spent in the medium. Thus, there is no increase in the Purcell factor, the cooperativity, or the effective $\chi^{(2)}$ nonlinearity in the lithium niobate.

In conclusion, we have demonstrated ultra-high-Q resonator modes in erbium-doped thin-film lithium niobate by engineering a low-loss slow light medium in fabricated microring cavities. We show agreement with theoretical models that account for the interference of resonator modes and the role of spin-bath induced dephasing. This platform has potential for a wide variety of quantum and classical photonic applications, particularly for overcoming the typical bandwidth mismatch between photonic and atomic systems. 

\section{Methods}
\subsection{Device design and fabrication}

We use grating couplers to couple light from angle-cut optical fiber into our waveguides. The grating couplers are designed such that they couple in and out only the transverse electric (TE) mode. The bus waveguide width is 800~nm, which allows us to effectively couple light into the 1200~nm width and $300~\mu\rm{m}$ diameter ring resonator.  The through and drop waveguides are positioned symmetrically on opposite sides of the ring resonator, each separated by a 800~nm gap  from the ring. Electrodes were placed $5~\mu\rm{m}$ away from the in-plane ring waveguides. 

We use a wafer with 600~nm of 0.5\% erbium-doped lithium niobate on insulator. We fabricate the devices using a two step electron-beam (e-beam) lithography process. First, we pattern the waveguides, rings, and grating couplers using ZEP e-beam resist. We then etch 300~nm using Argon based inductively coupled plasma reactive ion etching (ICP-RIE). The device is next cleaned of the remaining resist and redeposition of LN with a standard RCA clean. We anneal the sample in an O$_2$ environment at 500$^{\circ}$C for two hours. To fabricate the electrodes, we perform the second e-beam step using PMGI SF6 resist. We use electron beam deposition to deposit 15~nm of Cr and 300~nm of Au. We then clean and lift-off the deposited metals using a PG remover bath in a sonicated bath at 70$^{\circ}$C.

\subsection{Experimental Setup}
The devices were characterized in a 1.7~K closed cycle cryostat (Montana Instruments). The laser utilized is a Velocity 1628 tunable telecom laser whose polarization was controlled using a fiber polarization controller. A single acousto-optical modulator (AOM) (Aerodiode) is used to pulse the light and tune the frequency for probe scans. We used angle-cut fiber array v-grooves (OZ Optics) to couple light in and out of the grating couplers from the through port and the drop port (see Fig. S1). The sample is mounted on a 3-axis nanopositioner (AttoCube) to allow precise alignment of the fiber array with the grating couplers. The light collected from the device passes through another AOM, polarizer, and attenuator before being detected by a superconducting nanowire single photon detector (SNSPD) (PhotonSpot). The second AOM is used to shutter the SNSPD while high-intensity input pulses are used in burning or photoluminescence. 

\subsection{Linewidth narrowing using spectral hole burning}
Spectral hole burning was performed by applying a 50~$\mu$s burn pulse at a fixed frequency, followed by measuring the transmission of a probe pulse at the drop or through port. The probe pulse, with a duration of 5~$\mu$s and a weaker power of $P_{\text{probe}} = 40$~nW, was introduced 4~$\mu$s after the burn pulse. This process was repeated while maintaining a fixed burn laser frequency and sweeping the probe detuning from -20~MHz to 20~MHz. By recording the probe transmission at various detunings relative to the pump, we extracted the transmission of the narrowed resonance. The input power into the device was varied by changing the rf power supplied to the input AOM. It is important to note that the linewidth data used for analyzing power dependence and Fano resonance characteristics were acquired using a 5~ms burn pulse. This extended burn time ensured that the system reached a steady state, a critical assumption in our theoretical framework for deriving analytical solutions. In contrast, the 50~$\mu$s burn pulse resulted in an even narrower resonance feature, which was challenging to model due to the transient nature of the atomic population dynamics.

\subsection{Ring-down time and Slow light measurement}

The ring-down measurement was done by sending a square wave optical pulse of $1~\mu\rm{s}$ at a frequency where the narrowed resonance was burned. The pulse width was chosen to allow the the square pulse to fit spectrally within the narrowed resonance bandwidth. The trailing edge was fit with an exponential function $Ae^{-\tau t}$ where $\tau$ is the ring-down time.  The slow light measurement was done similarly by sending a Gaussian optical pulse of $1.5~\mu\rm{s}$ at the burn frequency. The pulse was fit with a gaussian and the center was compared with center of a reference pulse to give us $\tau_g$. In both these experiments, the reference was measured at a frequency of the bare cavity mode, which is a resonance away from the Er absorption. 

\subsection{Electro-optic tuning of Fano shape}
The electro-optic tuning was done by depositing gold electrodes around the ring resonator as described earlier. The two electrodes were wire bonded to two ports of a printed circuit board (see Fig. S5). The details of device architecture used in Fig. 3b are given in the Supplemental. A DC voltage was applied to produce a strong DC electric field across the waveguide, which changed the effective refractive index ($n_{\text{eff}}$) of the LN via its $\chi^{(2)}$ nonlinearity. Consequently, the resonance wavelength of the bare cavity shifts, as does the Fano \textit{q} parameter. \\

\noindent\textbf{Acknowledgments} \\
We acknowledge helpful discussions and support from Edo Waks, Uday Saha, Oğulcan E. \"{O}rsel, Yuqi Zhao, Christopher P. Anderson, and Kejie Fang. This work was supported by NSF QuIC-TAQS (Grant No. 2137642) and the Department of the Navy, Office of Naval Research.

\bibliography{Main_bib}% Produces the bibliography via BibTeX.

\end{document}

\renewcommand{\figurename}{Fig.}
\renewcommand{\thefigure}{S\arabic{figure}}
\renewcommand{\theequation}{S\arabic{equation}}

\maketitle
\section{Supplementary Theory}
\subsection{Strong driving of cavity coupled to ion ensemble}
We seek to understand the origins of the spectral features observed as a consequence of strongly driving a cavity weakly coupled to an ensemble of spectrally disordered saturable two level emitters. With this aim in mind, we eschew the phenomenological approach commonly adopted in photonics\,\cite{zhao2024cavity} in favor of a more fundamental approach relying on the Tavis-Cummings model. As we see in the ensuing analysis, such an approach is further necessitated by the power-dependent suppression of dephasing of the ions.  The Hamiltonian under the rotating wave approximation and in the driving frame of the exciting field for the cavity + emitters system is given as:
\begin{equation}
H = \Delta_{cl} a^\dagger a + \frac{1}{2} \sum_{j=1}^N \Delta_{jl} \sigma_j^z + \sum_{j=1}^N g_j \left(a^\dagger \sigma_j^- + \sigma_j^+ a\right)
- i\sqrt{\kappa_{\text{ext}}} A_l(a^\dagger - a).
\end{equation}
We have set $\hbar=1$. In the above Hamiltonian, $a$ is the cavity field operator, $\sigma_j^{\pm}$ is the raising/lowering operator for the $j$th ion and $\sigma_j^z$ is the Pauli-Z operator. The detuning of the cavity field resonant at frequency $\omega_c$ and that of the $j$th ion resonant at $\omega_j$, from the driving laser field has been denoted by $\Delta_{cl}$  and $\Delta_{jl}$  respectively. The $j$th ion couples to the cavity with strength $g_j$ and the cavity couples to the driving field with an external coupling rate $\kappa_{\text{ext}}$. The quantity $A_{l}$ is representative of the strength of the driving field and is related to its power $P_l$ by $A_l=\frac{P_l}{\hbar\omega_l}$, where $\omega_l$ is its frequency. The evolution of the cavity field and ionic operators under the action of the Hamiltonian is as follows, with the dissipation terms derived from modified Bloch equation formalism developed in\,\cite{geva1995relaxation,berman1985modified,schenzle1984microscopic} and summarized in \,\cite{yamanoi1986hole}:
\begin{align}
\dot{a} & = -(i\Delta_{cl} + \frac{\kappa}{2})a - i \sum_{j=1}^N g_j \sigma_j^- - \sqrt{\kappa_{\text{ext}}} A_l \\
\dot{\sigma}_j^- & = -\left(i\Delta_{jl} + \frac{\gamma_s}{2} +\gamma_d\frac{1+\Delta_{jl}^2\tau_c^2}{1+(|\Omega_j|^2+\Delta_{jl}^2)\tau_c^2}\right)\sigma_j^- + ig_j \sigma_j^z a +i\frac{\gamma_d\Omega_j\tau_c}{2}\frac{1-i\Delta_{jl}\tau_c}{1+(|\Omega_j|^2+\Delta_{jl}^2)\tau_c^2}\ \sigma_j^z \label{eqn:dynamiclowering}\\
\dot{\sigma}_j^z & = 2i g_j \left(a^\dagger \sigma_j^- - \sigma_j^+ a\right) - \gamma_s (1 + \sigma_j^z)
\end{align}
In the above set of equations, $\kappa$ is the decay rate of the bare cavity, $\gamma_s$ is the spontaneous emission decay rate, $\gamma_d$ is the dephasing rate in the absence of driving field or more realistically, in the presence of a weak field, $\Omega_j$ is the Rabi frequency associated with the $j$th ion in the cavity and $\tau_c$ is the bath correlation time. An erbium ion doped in lithium niobate experiences dephasing due to coupling with lithium and niobium nuclear spins, with a phonon bath via the direct phonon process and at high doping concentration, coupling with other erbium ions. At the magnetic field and temperature under which the measurements were carried out, we anticipate the lithium and erbium nuclear spin flips to be the predominant dephasing mechanism\,\cite{thiel2010optical}. Thus, the bath correlation time corresponds to the average spin flip time of the nuclear spins. Since the modification of the decoherence from the strong drive is not the primary focus of this work, we rely on existing Redfield type formalism which allows for an analytical description while adequately capturing the essential features of the phenomenon. The modified Bloch equations were developed in a semi-classical fashion with the field treated as a c-number\,\cite{geva1995relaxation}. In our case, a classical driving field excites the cavity field, $a$ which in turn excites the ions. Strictly speaking, the formalism would have to be revisited and further developed to incorporate the bosonic cavity field operator. This question has been addressed for a strongly driven cavity confined quantum dot which undergoes decoherence through phonon interactions\,\cite{wilson2002quantum}. To best of our knowledge, this problem has not been examined in the context of cavity confined rare-earth ions or defect centers coupled to a self-interacting bath. However, in our case, the application of the semi-classical modified Bloch equation in \eqref{eqn:dynamiclowering} is justified since the input power is high. The Rabi frequency, $\Omega_j$ depends on the cavity field as $\Omega_j=2g_j\langle a\rangle$. This makes the analytical treatment prohibitively difficult. We make a spate of approximations with the hope of gaining qualitative insight. We list the approximations below.
\begin{itemize}
    \item Since we are interested only in the steady state behavior, we parametrize the Rabi frequency in terms of the driving field $A_d$ by substituting the mean cavity field in the Rabi frequency expression with the steady state bare cavity field with the result $\Omega_j=-\frac{2g_j\sqrt{\kappa_{\text{ext}}}A_{\text{d}}}{i\Delta_{cl}+\kappa/2}$
    \item It has been demonstrated in \,\cite{yamanoi1986hole} that for very high Rabi frequencies when $|\Omega_j|\gg 1/\tau_c$, the dephasing can be adequately accounted even after dropping the last term in \eqref{eqn:dynamiclowering} and setting $\Delta_{jl}=0$ in the diagonal dephasing term. 
\end{itemize}
With the aid of these approximations, \eqref{eqn:dynamiclowering} is considerably simplified to: 
\begin{align}
    \dot{\sigma}_j^- & = -\left(i\Delta_{jl} + \gamma_{mj}\right)\sigma_j^- + ig_j \sigma_j^z a
\end{align}
Where $\gamma_{mj}$ is the modified decoherence rate of the $j$th ion and is to be interpreted as a shorthand for $\frac{\gamma_s}{2}+\frac{\gamma_d}
{1+|\Omega_j|^2\tau_c^2}$. We adopt the mean field approximation for the ionic and cavity field operators and evaluate their steady state by setting the time derivatives to zero. 
\begin{align}
    -(i\Delta_{cl} + \frac{\kappa}{2})\langle a\rangle - i \sum_{j=1}^N g_j\langle\sigma_j^-\rangle - \sqrt{\kappa_{\text{ext}}} A_l&=0\label{eqn:steadya}\\
    -\left(i\Delta_{jl} + \gamma_{mj}\right)\langle\sigma_j^-\rangle + ig_j \langle\sigma_j^z\rangle\langle a\rangle&=0\label{eqn:steadysigma}\\
    2i g_j \left(\langle a^\dagger\rangle\langle\sigma_j^-\rangle - \langle\sigma_j^+\rangle\langle a\rangle\right) - \gamma_s (1 + \langle\sigma_j^z\rangle)&=0\label{eqn:steadypauliz} 
\end{align}
We wish to express $\langle \sigma_j^z\rangle$ in terms of the mean cavity field, $\langle a\rangle$ which in turn is expressed in terms of the driving field and cavity parameters. We have chosen to solve for the steady state mean cavity field to include the influence of the high optical depth of ion ensemble on the bare cavity field. Our treatment is along the lines of \cite{lei2023many}.  Rearranging \eqref{eqn:steadysigma}, the polarization of the $j$th ion can be expressed as:
\begin{align}
    \langle\sigma_j^-\rangle=\frac{ig_j\langle\sigma_j^z\rangle\langle a\rangle}{i\Delta_{jl}+\gamma_{mj}}
\end{align}
We eliminate $\langle\sigma_j^-\rangle$ by substituting the above expression in \eqref{eqn:steadypauliz} and in \eqref{eqn:steadya} to yield:
\begin{align}
\langle\sigma_j^z\rangle &=-\frac{1}{1+\frac{4\gamma_{mj} g_j^2|\langle a\rangle|^2}{\gamma_s(\Delta_{jl}^2+\gamma_{mj}^2)}}\label{eqn:pauliz}\\
\langle a\rangle &= -\frac{\sqrt{\kappa_{\text{ext}}}A_l}{(i\Delta_{cl} +\frac{\kappa}{2})}\frac{}{\left[1-\sum_{j=1}^N\frac{g_j^2\langle\sigma_j^z\rangle}{(i\Delta_{jl}+\gamma_{mj})(i\Delta_{cl}+\frac{\kappa}{2})}\right]}\label{eqn:expandeda}
\end{align}
We bunch the terms in the denominator into a single quantity $I_l$ and evaluate it.
\begin{align}
    I_l\coloneqq & -\frac{1}{i\Delta_{cl}+\frac{\kappa}{2}}\sum_{j=1}^N\frac{g_j^2\langle\sigma_j^z\rangle}{i\Delta_{jl}+\gamma_{mj}}\label{eqn:Ildefinition}\\   
    &=\frac{1}{i\Delta_{cl}+\frac{\kappa}{2}}\sum_{j=1}^N\frac{g_j^2}{i\Delta_{jl}+\gamma_{mj}}\frac{1}{1+\frac{4\gamma_{mj}g_j^2|\langle a\rangle|^2}{\gamma_s(\Delta_{jl}^2+\gamma_{mj}^2)}}\label{eqn:expandedIl}
\end{align}
The ion-cavity coupling strength suffers from two sources of inhomogeneity: one arises from the position of a given ion in the cross-section of the ring waveguide and the other stems from the angular position of the ion in the ring with respect to the transition dipole moment axis. The ions located closer to the center of the cross-section experience a higher field and thus have a higher coupling strength than those at the edges. The second source of inhomogeneity is specific to our system: the electric field occupies the Transverse Electric (TE) mode in the waveguide and thus has a polarization in the plane of the microring. Since the transition dipole moment axis lies in plane, as the field propagates through the microring, the coupling strength varies as a function of the relative angle between the polarization and the moment axis. The spatial distribution of the ions is uniform and the position of the ion is uncorrelated with its transition frequency. As a result, both sources of coupling strength inhomogeneity and the spectral inhomogeneity can be treated mutually independent of one another. We denote the probability associated with the transverse inhomogeneity by $p(g)$ where $g$ is the coupling strength. The spectral inhomogeneous broadening was experimentally measured to be near Gaussian. However, the final results depend only on the spectral density of ions at a given frequency and not the shape of the distribution. For ease of analysis, we chose the spectral distribution profile to be Lorentzian with a full width half maximum that yields the correct local spectral density close to the center. The spectral density can thus be written as $\rho(\omega_j)=\frac{\Delta_{\text{inh}}}{2\pi}\frac{1}{(\omega_j-\omega_0)^2+(\Delta_{\text{inh}}/2)^2}$ where $\omega_0$ is the center of the broadening and $\Delta_{\text{inh}}$ is the FWHM. With the above considerations, we rewrite \eqref{eqn:expandedIl} by converting the summation to an integral over the ions.
\begin{align}
    I_l=\frac{1}{i\Delta_{cl}+\frac{\kappa}{2}}\int_g dg p(g)\int_0^{2\pi}\frac{d\theta}{2\pi}\int_{0}^{\infty} d\omega_j\rho(\omega_j)\frac{Ng^2\text{cos}^2(\theta)}{i(\omega_j-\omega_l)+\gamma_{m}}\frac{1}{1+ \frac{4\gamma_{m}g^2\text{cos}^2(\theta)|\langle a\rangle|^2}{\gamma_s\left((\omega_j-\omega_l)^2+\gamma_{m}^2\right)}} 
\end{align}
Where $\theta$ is the angle between the dipole moment axis and the polarization of the field. Since the coupling strength of the ion has been decoupled from its transition frequency, to avoid any notions of implicit correlation, we have dropped the $j$ index in $g_j$ as well as $\gamma_{mj}$ with $\gamma_m$ now denoting:
\begin{align}
    \gamma_m\coloneqq \frac{\gamma_s}{2}+\frac{\gamma_d}{1+\frac{4g^2\text{cos}^2(\theta)\kappa_{\text{ext}}A_d^2\tau_c^2}{\Delta_{cl}^2+(\kappa/2)^2}}
\end{align}
We first focus on the integral over frequency, $I_l^{(g,\theta)}$.
\begin{align}
 I_l^{(g,\theta)}= \int_{-\infty}^{\infty} d\omega_j\frac{\Delta_{\text{inh}}}{2\pi}\frac{Ng^2\text{cos}^2(\theta)}{(\omega_j-\omega_0)^2+(\Delta_{\text{inh}}/2)^2}\frac{\gamma_m-i(\omega_j-\omega_l)}{(\omega_j-\omega_l)^2+\gamma_{m}^2+ \frac{4\gamma_{m}}{\gamma_s}g^2\text{cos}^2(\theta)|\langle a\rangle|^2}  
\end{align}
 For analytical convenience, we have extended the lower limit of the integral to $-\infty$ since $\rho(\omega_j)\rightarrow 0$ when $\omega_j-\omega_0 \gg\Delta_{\text{inh}}$. The above integral is solved using the residue theorem to give:
 \begin{align}
     I_l^{(g, \theta)}=\frac{Ng|\text{cos}(\theta)|}{\Delta_{\text{inh}}}\frac{\sqrt{\gamma_s\gamma_m}}{\langle a\rangle}
 \end{align}
In arriving at the above result, we have assumed that $\omega_l-\omega_0 \ll \Delta_{\text{inh}}$. This approximation holds well since all the experiments have been carried out for the resonance mode of the cavity that nearly overlaps with the center of the inhomogeneous broadening. Additionally, we have assumed that broadening from the cavity field $\left(2\sqrt{\frac{\gamma_m}{\gamma_s}}g|\langle a\rangle|\right)$ is far greater than the ion's homogeneous broadening $\gamma_m$. We numerically verify this to be an excellent approximation even for the lowest input power considered. 
\begin{align}
    I_l=\frac{1}{i\Delta_{cl}+\frac{\kappa}{2}}\int_gdgp(g)\int_{0}^{2\pi}\frac{d\theta}{2\pi}\frac{Ng|\text{cos}(\theta)|}{\Delta_{\text{inh}}}\frac{\sqrt{\gamma_s\gamma_m}}{\langle a\rangle}\label{eqn:reducedil}
\end{align}
In the high-field limit, the integration over $\theta$ can be recast into the form $\int_0^{2\pi}\frac{d\theta}{2\pi}\sqrt{A\text{cos}^2(\theta)+B}$ where $A$ and $B$ are functions of the system parameters such that $A\gg B$. This integral can be expressed in terms of a standard elliptical integral of the second kind, $\frac{2}{\pi}\sqrt{A+B}\,E\left(\sqrt{\frac{A}{A+B}}\right)$ where $E(k)=\int_0^{\pi/2}d\theta\left(1-k^2\text{sin}^2(\theta)\right)$. Under the assumption $A\gg B$, we approximate $\sqrt{\frac{A}{A+B}}\approx 1$ and subsequently $E(1)=1$.  The result is equivalent to disregarding the $\theta$ dependence of $\gamma_m$ in \eqref{eqn:reducedil}.  
\begin{align}
    I_l=\frac{1}{i\Delta_{cl}+\frac{\kappa}{2}}\int_gdgp(g)\frac{2Ng}{\pi\Delta_{\text{inh}}}\frac{\sqrt{\gamma_s\gamma_m^\prime}}{\langle a\rangle}
\end{align}
where $\gamma_m^\prime\coloneqq \frac{\gamma_s}{2}+\frac{\gamma_d}{1+\frac{4g^2\kappa_{\text{ext}}A_d^2\tau_c^2}{\Delta_{cl}^2+(\kappa/2)^2}}$. The probability distribution, $p(g)$ follows from the transverse Gaussian profile of the TE mode in the waveguide and has log-uniform distribution. Due to the complicated dependence of the $\gamma_m^\prime$ on $g$, there is in general no simple solution. However, for a tightly confined field, due to the log-uniform nature of the distribution, we can replace $g$ by $\bar{g}$, the median value. We may approximate $I_l$ as
\begin{align}
    I_l=\frac{1}{i\Delta_{cl}+\frac{\kappa}{2}}\frac{2N\bar{g}}{\pi\Delta_{\text{inh}}}\frac{\sqrt{\gamma_s\bar{\gamma}_m}}{\langle a\rangle}
\end{align}
We have replaced the coupling strength in $\gamma_m^\prime$ by its mean value, $\bar{g}$ to give $\bar{\gamma}_m$. Substituting the above expression for $I_l$ in $\langle a\rangle=-\frac{\sqrt{\kappa_{\text{ext}}}A_d}{i\Delta_{cl}+\frac{\kappa}{2}}\frac{1}{1+I_l}$ \big(See \eqref{eqn:expandeda} and \eqref{eqn:Ildefinition}\big), we arrive at the following expression for the steady state cavity field. 
\begin{align}
    \langle a\rangle=-\frac{\sqrt{\kappa_{\text{ext}}}A_l}{i\Delta_{cl}+\frac{\kappa}{2}}\left(1+\frac{2N\bar{g}}{\pi\Delta_{\text{inh}}}\frac{\sqrt{\gamma_s\bar{\gamma}_m}}{\sqrt{\kappa_{\text{ext}}}A_l}\right)
\end{align}

\subsection{Weak probing of cavity coupled to ion ensemble}
Strongly driving the ions creates a transient spectral hole due to the driven erbium ions occupying the excited state. The millisecond long lifetime of the excited state and thereby, of the transient spectral hole far exceeds the short interval of time between the driving field and the application of the probe field. The probe field is weak enough that it doesn't perturb the population. The excited state population, when probed, thus remains nearly unchanged from its steady state value given by \eqref{eqn:pauliz}. The probe field at frequency $\omega$ is described by the following set of steady state equations:
\begin{align}
    -(i\Delta_{c\omega} + \frac{\kappa}{2})\langle a_p\rangle - i \frac{2\bar{g}}{\pi}\sum_{j=1}^N \langle{\sigma_j^-}_p\rangle - \sqrt{\kappa_{\text{ext}}} A_p&=0\label{eqn:steadyaprobe}\\
    -\left(i\Delta_{j\omega} + \gamma\right)\langle{\sigma_j^-}_p\rangle + i\frac{2\bar{g}}{\pi} \langle\sigma_j^z\rangle\langle a_p\rangle&=0\label{eqn:steadysigmaprobe} 
\end{align}
We have adopted the mean-field approximation as we did earlier. The quantities $A_p$, $a_p$ and ${\sigma_j^-}_p$ denote the input probe field, the cavity field under probe field excitation and the polarization of the $j$th ion in response to the probe field respectively. $\Delta_{c\omega}$ and $\Delta_{j\omega}$ are the detunings of the cavity and $j$ th ion from the probe frequency. Since the probe field is weak, the decoherence rate $(\gamma=\gamma_s/2+\gamma_d)$ can be treated as being independent of the probe field strength. Following the results of the preceding section, we have replaced $g_j$ and $\gamma_{mj}$ with $\frac{2g}{\pi}$ and $\bar{\gamma}_m$ respectively in anticipation of the outcome of the summation over the transverse and angular inhomogeneity. Consequently, in the ensuing analysis, we consider only the spectral inhomogeneous broadening. We eliminate $\langle{\sigma_j^-}_p\rangle$ by rearranging \eqref{eqn:steadysigmaprobe} and substituting the resulting expression in \eqref{eqn:steadyaprobe} to yield: 
\begin{align}
\langle{\sigma_j^-}_p\rangle&=\frac{2i\bar{g}\langle\sigma_j^z\rangle\langle a_p\rangle}{\pi\left(i\Delta_{j\omega} + \gamma\right)}\\
    \langle a_p\rangle&=-\frac{\sqrt{\kappa_{\text{ext}}}A_p}{i\Delta_{c\omega}+\frac{\kappa}{2}}\,\frac{1}{1-\frac{4\bar{g}^2}{\pi^2\left(i\Delta_{c\omega}+\frac{\kappa}{2}\right)}\sum_{j=1}^N\frac{\langle \sigma_j^z\rangle}{i\Delta_{j\omega}+\gamma}}
\end{align}
The cavity field under probe field excitation is a function of $\langle\sigma_j^z\rangle$ and thus maps the steady state population achieved through strong driving. Adopting the same approach as in the previous section, we evaluate the terms in the denominator. 
\begin{align}
I_{\omega}\coloneq\frac{4N\bar{g}^2}{\pi^2\left(i\Delta_{c\omega}+\frac{\kappa}{2}\right)}\int_{-\infty}^\infty d\omega_j \frac{\rho(\omega_j)}{i(\omega_j-\omega)+\gamma}\frac{1}{1+\frac{16\bar{\gamma}_{m} \bar{g}^2|\langle a\rangle|^2}{\pi^2\gamma_s\left((\omega_j-\omega_l)^2+\bar{\gamma}_{m}^2\right)}}
\end{align}
As usual, we have converted the summation to an integral by reintroducing the spectral density function and have substituted the expression for $\langle\sigma_j^z\rangle$ from \eqref{eqn:pauliz}. We recognize the quantity $\frac{16\bar{g}^2|\langle a\rangle|^2}{\pi^2\gamma_s\bar{\gamma}_m}$ to be the dimensionless saturation parameter and in the interest of brevity, replace the bloated group of terms with $s$
\begin{align}
   I_{\omega}\coloneq\frac{4N\bar{g}^2}{\pi^2\left(i\Delta_{c\omega}+\frac{\kappa}{2}\right)}\int_{-\infty}^\infty d\omega_j\frac{\Delta_{\text{inh}}}{2\pi}\frac{1}{(\omega_j-\omega_0)^2+(\Delta_{\text{inh}}/2)^2}\frac{1}{i(\omega_j-\omega)+\gamma} \frac{(\omega_j-\omega_l)^2+\bar{\gamma}_m^2}{(\omega_j-\omega_l)^2+\bar{\gamma}_m^2+s\bar{\gamma}_m^2} 
\end{align}
We analytically solve the above integral using the residue theorem and in the process assume that the inhomogeneous broadening is much larger than any detunings since we are probing a very narrow region of frequency spanning 40 MHz around the drive frequency. We have also assumed that the inhomogeneous broadening is far greater than the power broadening and used approximations to that effect. Its validity is verified for the largest input power that we used: $21 ~ \mu\text{W}$ which results in power broadening $(\sqrt{s}\bar{\gamma}_m/\pi)$ of  $0.15 \text{ GHz}$ while the inhomogeneous broadening $(\Delta_{\text{inh}}/2\pi)$ is $158 \text{ GHz}$. Clearly, the approximation is warranted.   
\begin{align}
 I_{\omega}=\frac{8N\bar{g}^2}{\pi^2\Delta_{\text{inh}}\left(i\Delta_{c\omega}+\frac{\kappa}{2}\right)}\left\{\frac{\omega-\omega_l}{\omega-\omega_l+i\sqrt{s}\bar{\gamma}_m}\right\}
\end{align}
With this result, we obtain the expression for the cavity field under probe field excitation. 
\begin{align}
\langle a_p\rangle&=-\frac{\sqrt{\kappa_{\text{ext}}}A_p}{i\Delta_{c\omega}+\frac{\kappa}{2}}\,\frac{1}{1+I_{\omega}}    
\end{align}
In this form, the expression inspires little qualitative insight into the observed spectral profiles of the probe field. We devote the next section for this purpose. 

\subsection{Fano Profiles \& Linewidth Narrowing}
\subsubsection{Fano Profile}
Considering the results in the preceding section, we can express the cavity field as: 
\begin{align}
\langle{a_p\rangle}=-i\sqrt{\kappa_{\text{ext}}}A_{p}\frac{\omega_l-\omega-i\sqrt{s}\bar{\gamma}_m}{\left(\omega_c-\omega-i\left(\frac{\kappa}{2}+\frac{8N\bar{g}^2}{\pi^2\Delta_{\text{inh}}}\right)\right)\left(\omega_l-\omega-i\sqrt{s}\bar{\gamma}_m\right)+\frac{8N\bar{g}^2}{\pi^2\Delta_{\text{inh}}}\sqrt{s}\bar{\gamma}_m}\label{eqn:fano1}
\end{align}
In this form, the cavity field spectrum resembles the spectrum of a driven mode of a resonator coupled to a second resonator\,\cite{limonov2017fano}. The quantity, $\frac{\sqrt{s}\bar{\gamma}_m}{\pi}$ is the power broadened spectral hole width ($\frac{\Delta_{\text{hole}}}{2\pi}$) and $\frac{16N\bar{g}^2}{\pi^2\Delta_{\text{inh}}}$ is the cavity loss rate due to erbium ion absorption ($\kappa_{\text{Er}}$).
\begin{align}
   \Delta_{\text{hole}}\equiv & 2\sqrt{s}\bar{\gamma}_m\\
   \kappa_{\text{Er}}\equiv & \frac{16N\bar{g}^2}{\pi^2\Delta_{\text{inh}}}
\end{align}
The broader resonator with the FWHM $(\frac{\kappa_{\text{cont}}}{2\pi})$ of $\frac{\kappa}{2\pi}+\frac{\kappa_{\text{Er}}}{2\pi}$ is the resonance of the cavity + ion ensemble system in the absence of the strong drive i.e. $s=0$. The transient hole of width $\frac{\Delta_{\text{hole}}}{2\pi}$ generated by the strong drive acts like a narrow defect resonator of the same width $(\frac{\kappa_{\text{def}}}{2\pi})$. The interference of the defect resonator mode with the broad resonance of the cavity + ion ensemble creates Fano type profiles. The broad resonator mode thus behaves like a quasi-continuum. 

The effective coupling $(\nu)$ between the quasi-continuum and the defect resonance mode is $\frac{\sqrt{\kappa_{\text{Er}}\Delta_{\text{hole}}}}{2}$ and this coupling mutually shifts the resonance frequency of both modes in opposite directions to yield shifted resonance frequencies $\tilde{\omega}_c$ and $\tilde{\omega}_l$ 
\begin{align}
    \tilde{\omega}_c=\omega_c-\frac{\sqrt{\kappa_{\text{Er}}\Delta_{\text{hole}}}}{2}\\
\tilde{\omega}_l=\omega_l+\frac{\sqrt{\kappa_{\text{Er}}\Delta_{\text{hole}}}}{2}
\end{align}
We denote the detuning from $\tilde{\omega}_l$ by $\epsilon$ such that $\epsilon\equiv\omega-\tilde{\omega}_l$. Adopting the approach followed in \cite{iizawa2021quantum}, we rearrange \eqref{eqn:fano1} to resemble the standard Fano lineshape and deduce the Fano parameter, $q$.
\begin{align}
    \langle a_p\rangle=\frac{-i\sqrt{\kappa_{\text{ext}}}A_p}{\omega_c-\omega_l}\left\{\frac{\frac{\epsilon (\omega_c-\omega_l)^2}{\frac{\sqrt{\kappa_{\text{Er}\Delta_{\text{hole}}}}}{2\pi}\left(\frac{\kappa}{2}+\frac{\kappa_{\text{Er}}}{2}\right)}-\frac{\omega_c-\omega_l}{\frac{\kappa}{2}+\frac{\kappa_{\text{Er}}}{2}}\left(1+i\frac{\omega_c-\omega_l}{\frac{\kappa_{\text{Er}}}{2}}\right)}{  \frac{\epsilon (\omega_c-\omega_l)^2}{\frac{\sqrt{\kappa_{\text{Er}\Delta_{\text{hole}}}}}{2\pi}\left(\frac{\kappa}{2}+\frac{\kappa_{\text{Er}}}{2}\right)}-i}\right\}\equiv\frac{-i\sqrt{\kappa_{\text{ext}}}A_p}{\omega_c-\omega_l}\frac{\Omega'+q}{\Omega'-i}
\end{align}
 In deriving the above expression, we have assumed that we are probing a very narrow region of frequency around the defect resonance frequency and made approximations to that effect. We have recognized the first term in the numerator and denominator to be the renormalized detuning, $\Omega'$ and the second term in the numerator to be the complex Fano parameter, $q$.
 \begin{align}
 \Omega'\equiv  \frac{\epsilon (\omega_c-\omega_l)^2}{\frac{\sqrt{\kappa_{\text{Er}\Delta_{\text{hole}}}}}{2\pi}\left(\frac{\kappa}{2}+\frac{\kappa_{\text{Er}}}{2}\right)}\\
 q\equiv \frac{\omega_c-\omega_l}{\frac{\kappa}{2}+\frac{\kappa_{\text{Er}}}{2}}\left(1+i\frac{\omega_c-\omega_l}{\frac{\kappa_{\text{Er}}}{2}}\right)
\end{align}

\subsubsection{Linewidth Narrowing}
Having recognized the observed probe field spectrum as arising effectively from the coupling between a narrow defect resonator and the quasi-continuum, we direct our attention to the specific case of $q=0$. This is achieved by driving the cavity at its resonant frequency such that $\omega_l-\omega_c=0$. On probing, a quasi-Lorentzian dip is observed in the reflection spectrum and a Lorentzian peak is observed in the transmission spectrum. Setting the cavity detuning to zero in \eqref{eqn:fano1}, we compute the linewidth to be $\frac{\Delta_{\text{hole}}}{2}\left(1-\frac{C}{1+C}\right)$
\begin{align}
 \langle a_p\rangle =\frac{\sqrt{\kappa_{\text{ext}}}A_p}{\frac{\kappa}{2}(1+C)}\frac{\omega_l-\omega-i\Delta_{\text{hole}}/2}{\omega_l-\omega-i\frac{\Delta_{\text{hole}}}{2}\left(1-\frac{C}{1+C}\right)}\label{eqn:fanolorentz}  
\end{align}
$C$ is the ensemble co-operativity defined as $\frac{\kappa_{\text{Er}}}{\kappa}$. From the coupled resonator point of view, the quantity $\frac{C}{1+C}$ is the ratio of the defect resonator-quasi-continuum coupling strength to the decay rates in the defect and the continuum. We thus identify it to be the co-operativity of the defect resonator.
\begin{align}
    C_{\text{def}}\equiv\frac{4\nu^2}{\kappa_{\text{def}}\kappa_{\text{cont}}}=\frac{C}{1+C}
\end{align}
When the ensemble co-operativity is low i.e. $C\ll 1$, the observed linewidth is simply the linewidth of the defect resonance mode. This is unsurprising since the defect co-operativity is correspondingly low and the excitation remains largely limited to the quasi-continuum. For large ensemble co-operativity, as is the case in our system, the defect co-operativity approaches unity. This implies that the losses in the system are balanced by the coupling of energy into the defect resonator mode. Subsequently, upto first order, the observed linewidth is narrowed to $\frac{\Delta_{\text{hole}}}{2\pi C}$.

We show that the obtained results are consistent with the more phenomenological interpretation adopted in \cite{sabooni2013spectral} where the narrowing is attributed to the reduction in group velocity of light within the spectral hole. Expanding the ensemble co-operativity, we rewrite the observed linewidth due to the slow light as $ \frac{\kappa_{SL}}{2\pi} = \frac{\Delta_{\text{hole}}}{\kappa_{\text{Er}}}\frac{\kappa}{2\pi}$. The ratio of the width of the transient spectral hole to the absorption rate determines the group velocity of light $(v_{g})$ in the spectral hole. With this interpretation, the observed linewidth  is now $\frac{nv_g}{c}\frac{\kappa}{2\pi}$. Thus, within the spectral hole, the linewidth of the bare cavity is scaled down by the group velocity.     

So far, we have focused on the amplitude of the spectrum. The coupling of the defect resonator to the quasi-continuum also introduces a frequency dependent phase, $\Phi(\omega)$. The phase of the cavity field in \eqref{eqn:fanolorentz} is:
\begin{align}
    \Phi(\omega)=\arctan{\frac{\Delta_{\text{hole}}}{2(\omega-\omega_l)}}-\arctan{\frac{\Delta_{\text{hole}}/C}{2(\omega-\omega_l)}}
\end{align}
The group delay $(\tau_D)$ is given by the first derivative of the phase with respect to frequency while the second derivative gives the group delay dispersion $\frac{\text{d}\tau_D}{\text{d}\omega}$. At the center of the hole group delay is finite and significant while the group delay dispersion is zero. The dispersion becomes significant at the edges of the hole.
\begin{align}
    \tau_D=\frac{\text{d}\Phi}{\text{d}\omega}\bigg\vert_{\omega_l}=\frac{2C}{\Delta_{\text{hole}}}
\end{align}

\section{Supplementary Experiments}
\begin{figure}[H]
\includegraphics[width=\linewidth,keepaspectratio]{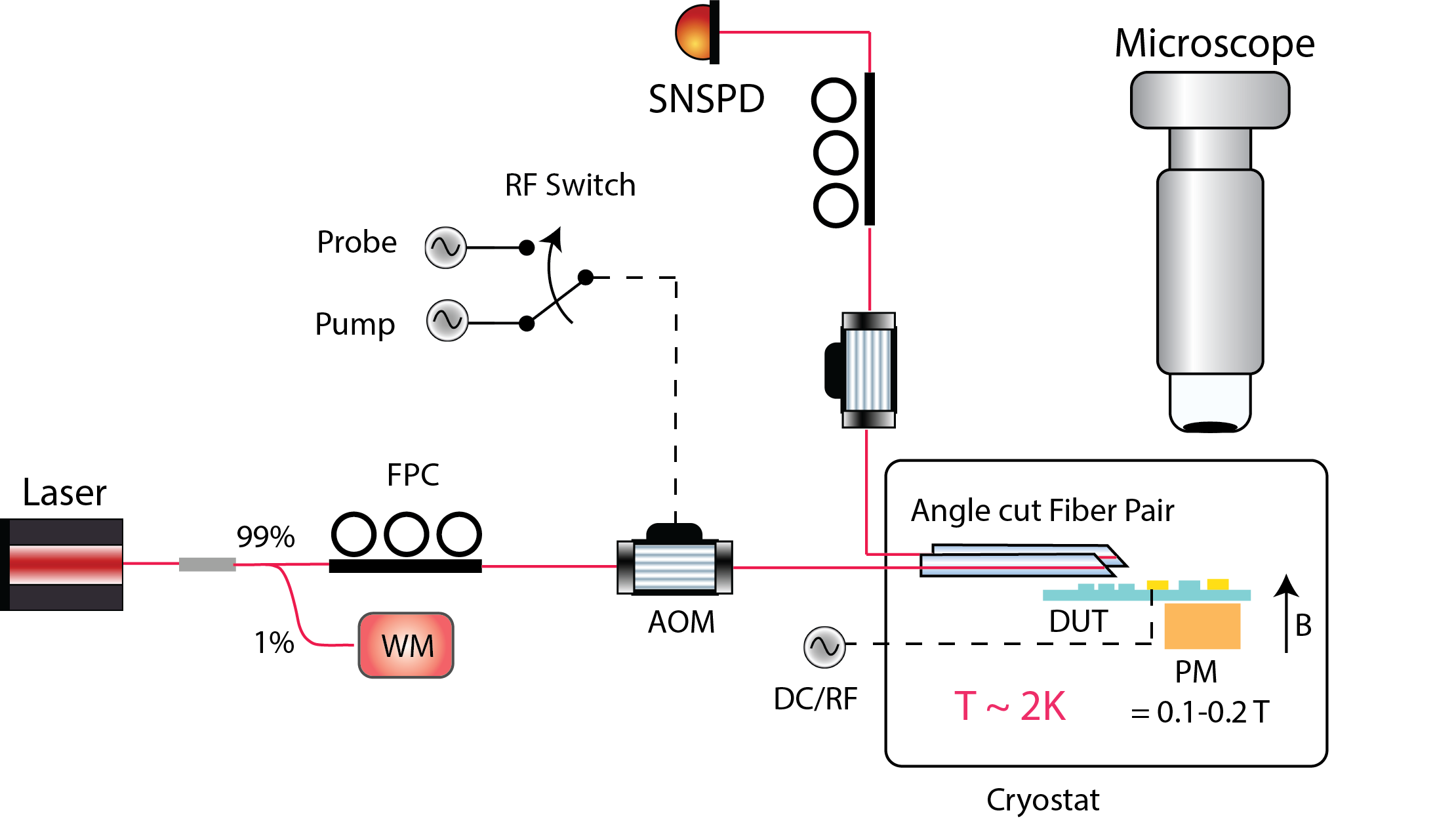}
\caption{\label{fig:Setup}Schematic of the measurement setup - AOM : acousto-optic modulator, FPC - Fiber polarization control, SNSPD : Superconducting nanowire single photon detector, WM : Wavemeter, PM : Permanent Magnet, DUT : Device Under Test, DC/RF- Function generator which produces DC and RF signal }
\end{figure}

\subsection{Extracted Parameters}
\begin{table}[H]
    \centering
    \begin{tabular}{l c l}
        \toprule
        \textbf{Parameter (Symbol)} & \textbf{Value}\\
        \midrule
        Spontaneous Decay Rate ($\gamma_s$) & $2\pi \times (56  \pm 1) \text{ Hz}$ \\
        Decoherence Rate ($\gamma$) & $2\pi \times ( 123 \pm 19) \text{ kHz}$  \\
        Single Ion-Cavity Coupling Rate ($g$) & $2\pi \times (362 \pm18)$ kHz \\
        Absorption Rate ($\kappa_{\text{Er}}$) &   $ \approx  2\pi \times 16 \text{ GHz} $ \\
        Cavity Internal Coupling Rate ($\kappa_{\text{ext}}$) & $2\pi \times (324 \pm 17)$  MHz  \\
        Cavity External Coupling Rate ($\kappa_i$) & $2\pi \times (471 \pm 54)$ MHz \\
        Cavity Energy Decay Rate ($\kappa$)  & $2\pi \times (1.12 \pm 0.04)  \text{ GHz}$ \\
        \bottomrule
    \end{tabular}
    \caption{Extracted Model parameters from independent measurements described in the following section}
    \label{tab:parameters}
\end{table}

\subsection{\textit{g} measurement and Purcell Factor}
\begin{figure}[H]
\includegraphics[width=\linewidth,keepaspectratio]{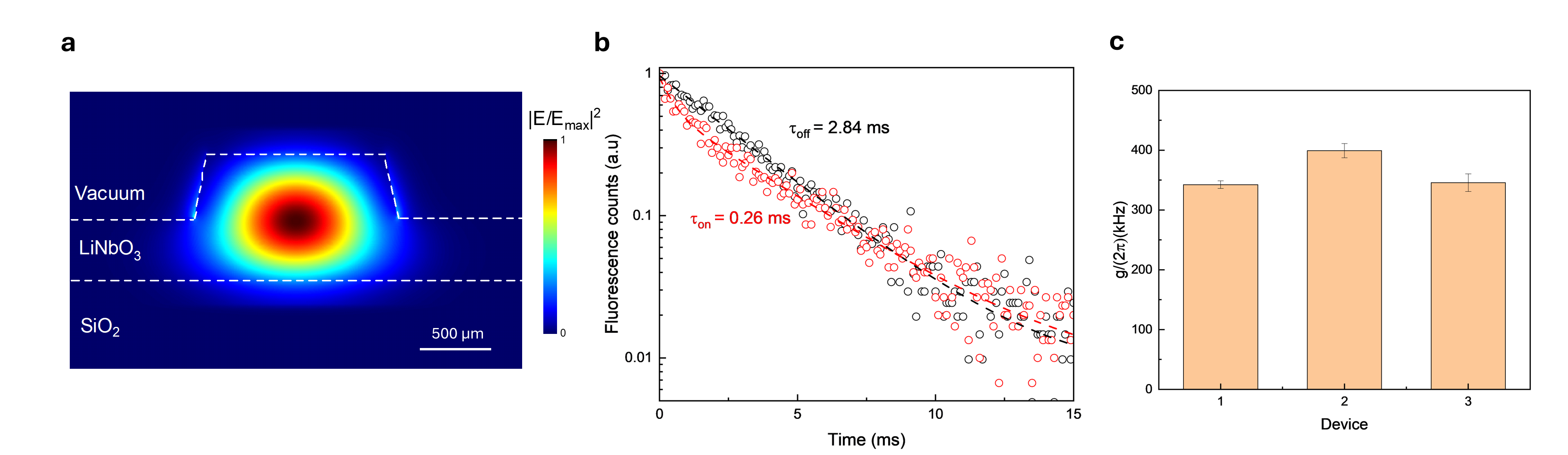}
\caption{\label{fig:Purcell} (a) Normalized $\lvert E \rvert^2$ field distribution in the ring waveguide, demonstrating the spatial confinement of the optical mode within the resonator.
(b) Measured fluorescence lifetime of Er ions when coupled to a resonator mode near the edge of the inhomogeneous broadening (red), compared to the off-resonant case where emission is primarily collected from the bus waveguide (black). (c) The estimated g for three different devices with $\kappa =2\pi \times \{2, 1.12, 0.99\} \text{ GHz}$. The variation in $\kappa$ was obtained by using devices with different separations from the bus waveguide.
}
\end{figure}

In the weak cavity coupling regime, the cavity provides an additional channel for the spontaneous emission of Er ions, described by the expression \cite{wang2020incorporation}
\begin{equation}
\label{eq:Purcell}
\gamma_P = \kappa \frac{g^2}{\kappa^2 / 4 + \Delta_{cl}^2} +\gamma_s
\end{equation}
where the $\gamma_P$ is the Purcell-enhanced rate of spontaneous emission in the cavity. To determine $\gamma_s$, we measured the spontaneous emission rate off a resonant mode, where emission occurs primarily through the bus waveguide. The fluorescence decay follows a single exponential function, $Ae^{-\gamma_s t}$ (Fig. \ref{fig:Purcell}b). By fitting this exponential to the measured fluorescence emission, following a 1 ms pump, we obtain $\gamma_s =2\pi \times (56  \pm 1) \text{ Hz}$, corresponding to a spontaneous emission lifetime of \(T_1 = 1/\gamma_s = 2.84 \pm 0.04\) ms.

Next, we performed the same measurement with the laser frequency tuned to zero detuning with the cavity resonance (\(\Delta_{cl} = 0\)). To minimize absorption effects while determining $g$, which is the single-photon coupling rate for the bare cavity, we positioned the laser at a resonator at the edge of the inhomogeneous broadening. In this case, fluorescence emission originates from both the bus waveguide and the ring resonator, leading to a decay described by $A_1e^{-\gamma_s t} +A_2e^{-\gamma_P t}$. By fixing \(\gamma_s\) from the previous measurement, we extracted \(\gamma_P = 2 \pi \times (0.624  \pm 0.03)\) kHz (Fig. \ref{fig:Purcell}b). The corresponding Purcell factor is calculated as $P = \gamma_P/\gamma_s -1 = 10.1 \pm 0.5 $. 

Using these values in Eq. \ref{eq:Purcell}, we calculated the coupling strength \( g = 2\pi \times (399 \pm 11)\) kHz. To further validate our analysis, we repeated the measurement on three different devices with varying \(\kappa\), obtaining an average \( g = 2\pi \times (362 \pm18\)) kHz  .

\subsection{Unmodified $\gamma$ estimation} 
\begin{figure}[H]
\includegraphics[width=\linewidth,keepaspectratio]{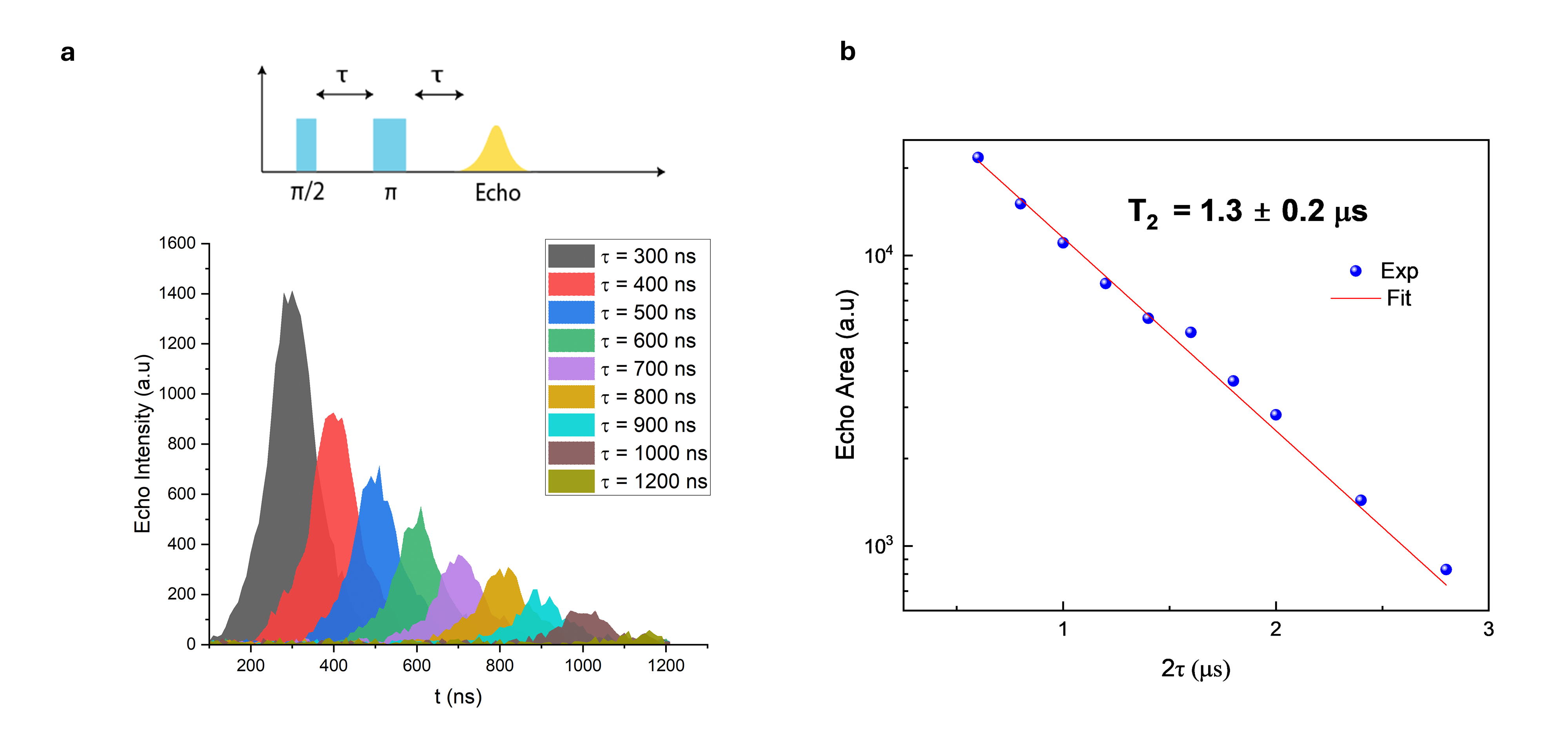}

\caption{\label{fig:PhotonEcho} The optical decoherence rate was measured using photon echo (a) showcases the photon echo intensity measured for different pulse separation between the $\pi/2 $ and $\pi $ described as $\tau$ (b) the echo area measured from part (a) was plotted against $2\tau$ and fit using \ref{eqn:Photon Echo} to extract the $T_2$}
\end{figure}

Two-photon echo is a standard technique to extract the phase-coherence time for atomic ensembles. Two short pulses of 100 ns and 200 ns were separated by a time delay $\tau$, which resulted in a coherent echo signal after time $\tau$ (Fig. \ref{fig:PhotonEcho}). By fitting the echo peak intensity $I(\tau)$ to the expression, we were able to extract the decoherence rate $\gamma = 1/T_2  = 2\pi \times ( 123 \pm 19) \text{ kHz}$.   

\begin{equation} 
\label{eqn:Photon Echo}
I(\tau) = I_0 \exp \left[ -2 \left( \frac{2\tau}{T_2} \right)^x \right]
\end{equation}
Where $x > 1$ is a measure of the spectral diffusion. The homogenous linewidth can be further estimated to be  $\Gamma_h = 1/(\pi T_2) = 245 \pm 38 \text{ kHz} $

\subsection{Inhomogeneous broadening and absorption coefficient ($\alpha_{\text{Er}}$)}
The absorption coefficient in Er:LN devices, was extracted using a waveguide configuration with dimensions identical to those of the ring resonator. In the ring, the optical mode (transverse electric) samples both in-plane orientations of the Er-induced absorption due to the circular geometry. To emulate this behavior in a linear structure and accurately determine the absorption rate, we fabricated a series of spiral waveguides with an increasing number of turns. Each additional turn increases the waveguide length, enabling the electric field to sample a full 360° range of crystallographic orientations relative to the Er absorption axis, similar to the ring. By measuring transmission of a weak probe power ($<$pW) sent through the waveguides and extracting the absorption coefficient as shown in Fig. \ref{fig:Abs}a.

\subsection{Extracting coupling and loss coefficients ($\kappa_{\text{ext}}, \kappa_i, \kappa_{\text{Er}}$)} \label{Rates}
To extract the coupling loss rate (\(\kappa_{\text{ext}}\)) and the intrinsic loss rate (\(\kappa_i\)), we move away from the absorption spectrum and characterize the transmission of the bare cavity mode (Fig. 1). The through-port transmission exhibits a Lorentzian dependence, given by:

\[
T =  \left(1 - \frac{\kappa_{\text{ext}}}{\mathrm{i} \Delta_{cl} + \frac{\kappa}{2}}\right)^2
\]

where the total loss rate is defined as \(\kappa = 2\kappa_{\text{ext}} + \kappa_i\). This total loss accounts for propagation losses (\(\kappa_i\)) and coupling losses due to the two bus waveguides (\(\kappa_{\text{ext}}\)). By fitting this expression to the transmission data, we obtain \(\kappa_{\text{ext}} = 2\pi \times (324 \pm 17) \) MHz and \(\kappa_i = 2\pi \times (471 \pm 54)\) MHz.

\begin{figure}[H]
\includegraphics[width=\linewidth,keepaspectratio]{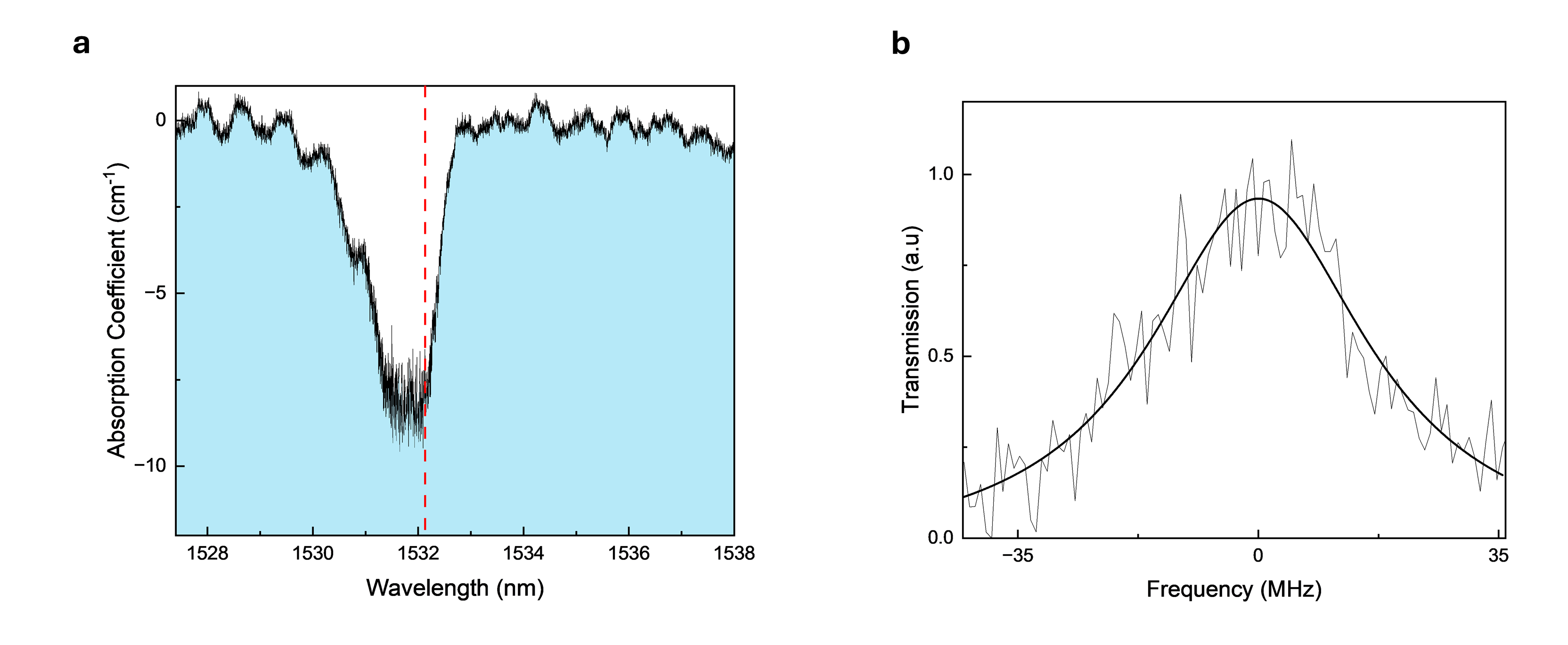}
\caption{\label{fig:Abs} (a) The measured absorption coefficient of $\text{Er}^{3+}$ in LN, demonstrating an inhomogeneous broadening of 158 GHz, the dotted line represents the wavelength at which the spectral hole burning experiments were conducted (b) A spectral hole burning in a waveguide showcasing a transparency window.}
\end{figure}

Directly measuring the ion-induced loss rate (\(\kappa_{\text{Er}}\)) was challenging near the peak of the inhomogeneous broadening. The strong absorption significantly broadened the cavity resonance and reduced the extinction ratio over the varying background of the inhomogeneous broadening. Moreover, the excess nonlinearity induced by Er made it challenging to avoid dragging or narrowing of the resonance feature while scanning without saturating the emitter. Instead, we estimate \(\kappa_{\text{Er}}\)  using the attenuation coefficient extracted earlier, \(\alpha_{\text{Er}}\ \approx 8 ~ \text{cm}^{-1}\) at the frequency where the measurement was made (Fig. \ref{fig:Abs}a). The total loss rate can be expressed by \cite{mckinnon2009extracting} :

\begin{equation}
\label{eq:Rates}
    \kappa = \frac{2c \cdot \cos^{-1} \left( \frac{2ar}{1 + (ar)^2} \right)}{n_g L}
\end{equation}

where \(a\) is the single-pass amplitude transmission, \(r\) is the self-coupling coefficient, \(L\) is the cavity length, and \(n_g\) is the group index of the waveguide mode without any spectral hole burning (Fig. \ref{fig:Purcell}a). The power attenuation in a single round trip is given by \(a^2 = e^{-\alpha L}\), where \(\alpha = \alpha_p + \alpha_{\text{Er}}\) represents the total attenuation coefficient, consisting of propagation losses (\(\alpha_p\)) and absorption losses (\(\alpha_{\text{Er}}\)). In the case of an bare cavity mode (\(\alpha_{\text{Er}} = 0\)), we extract the product \(ar = 0.97\) from the \(\kappa\) obtained previously.

When additional attenuation due to erbium absorption (\(\alpha_{\text{Er}}\)) is introduced, the single-pass amplitude transmission modifies to \(a_{\text{Er}} = e^{-\alpha_{\text{Er}}L/2}a\). Using this modified product \(a_{\text{Er}}r\) in Equation Eq. \ref{eq:Rates}, we extract $\kappa_{\text{Er}} \approx 2\pi \times 16\ \text{GHz}$.

\subsection{Spectral Hole burning in Waveguide}
Spectral hole burning in the waveguide was performed using a short waveguide segment with an effective optical depth (OD) of 0.38. A 5~ms burn pulse was followed by an 80~MHz probe scan conducted over a 10~\textmu s window. To achieve this wide scan range, we used two AOMs in series before coupling the light into the device. The remainder of the setup was identical to the setup described in Figure \ref{fig:Setup}. 

As a representative example (Fig. \ref{fig:Abs}b), we used a pump power of 120~nW, corresponding to a Rabi frequency of $2\pi \times 780$~kHz. This is lower than the Rabi frequency observed in the ring resonator ($ \sim 2\pi \times 10 \text{ MHz} $) at the same power. Nonetheless, we observed broader hole with $\Delta_{\text{hole}} = 2\pi \times 39 \text{ MHz}$ (Fig. \ref{fig:Abs}b), significantly wider than those in the resonator, with a linewidth of $\kappa_{\text{SL}} \approx 2\pi \times 3$~MHz. This observation confirms that the narrow features observed in the resonator are not solely due to reduced absorption. If they were, we would expect the resonator linewidth to be comparable to the spectral hole width, i.e., $\kappa_{\text{SL}} \sim \Delta_{\text{hole}}$. Instead, we measure $\kappa_{\text{SL}} < \Delta_{\text{hole}}$, indicating that the resonator features are primarily governed by dispersion effects rather than reduced absorption alone.

\subsection{Electro-Optic tuning}
\begin{figure}[H]
\includegraphics[width=\linewidth,keepaspectratio]{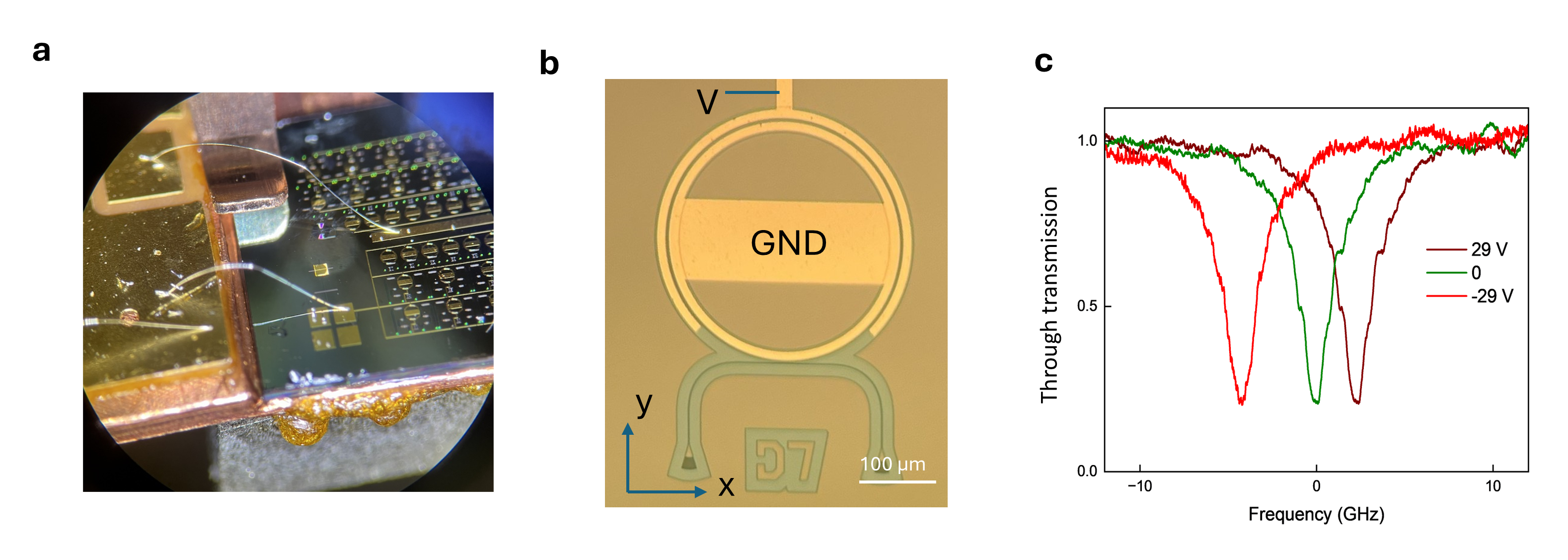}
\caption{\label{fig:EOtuning} (a) Photograph of the fabricated devices wire bonded to a PCB board inside our setup  (b) Representative of the device used in the Fano tuning experiment in Fig. 3b. The EO tuning of the bare cavity resonance when a DC voltage is applied}
\end{figure}

The electro-optic tuning results showcased in Fig. 3b, were done in a single-port device configuration (Fig. \ref{fig:EOtuning}b) to achieve optimal performance. This design enabled access to the stronger second-order nonlinear susceptibility ($\chi^{(2)}$) along the y-axis of lithium niobate (LN), characterized by an electro-optic coefficient of $r_{22} = 7$ pm/V. Additionally, it provided improved overlap between the applied electric field and the optical mode of the ring resonator. The DC voltage required for tuning was supplied via a printed circuit board (PCB) mounted adjacent to the chip, allowing for direct wire bonding to the on-chip electrodes. As shown in Fig. \ref{fig:EOtuning}c, applying a DC voltage across the electrodes induced a phase shift in the region of the ring resonator, resulting in a shift of the resonance frequency. We measured an electro-optic tuning rate of 0.12 GHz/V.

The Stark shift in erbium-doped LN is predominantly observed when the electric field is aligned along the z-axis, with a measured sensitivity of approximately 25 kHz/(Vcm$^{-1}$) \cite{yang2023controlling}. In our device, the electric field is oriented along the y-axis—perpendicular to the z-axis—so we expect the contribution from the Stark effect to be minimal.

\bibliography{Supplementary_bib}
\bibliographystyle{naturemag}